\documentclass[aps,prc,preprint,amsmath,amssymb,showpacs,preprintnumbers,superscriptaddress]{revtex4-1}
\usepackage{CJK}
\usepackage{graphicx}
\usepackage{dcolumn}
\usepackage{bm}
\usepackage{mathrsfs}   
\usepackage{color,xcolor}  
\usepackage{multirow}  
\usepackage{booktabs} 

\usepackage{hyperref}

\usepackage{mathtools}    

\newcommand{\la}{\langle}
\newcommand{\ra}{\rangle}

\newcommand{\bbm}{\begin{bmatrix}}
\newcommand{\ebm}{\end{bmatrix}}
\newcommand{\bBm}{\begin{Bmatrix}}
\newcommand{\eBm}{\end{Bmatrix}}
\newcommand{\bpm}{\begin{pmatrix}}
\newcommand{\epm}{\end{pmatrix}}

\begin{document}


\title{Microscopic optical potential from the relativistic Brueckner-Hartree-Fock theory: Proton-nucleus scattering}

\author{Pianpian Qin}
\affiliation{Department of Physics and Chongqing Key Laboratory for Strongly Coupled Physics, Chongqing University, Chongqing 401331, China}

\author{Sibo Wang}
\email{sbwang@cqu.edu.cn}
\affiliation{Department of Physics, Chongqing Key Laboratory for Strongly Coupled Physics, Chongqing University, Chongqing 401331, China}

\author{Hui Tong}
\affiliation{Helmholtz-Institut f$\ddot{u}$r Strahlen- und Kernphysik and Bethe Center for Theoretical Physics, Universit$\ddot{a}$t Bonn, D-53115 Bonn, Germany}

\author{Qiang Zhao}
\affiliation{Center for Exotic Nuclear Studies, Institute for Basic Science, \\
Daejeon 34126, Korea}

\author{Chencan Wang}
\affiliation{School of Physics and astronomy, Sun Yat-Sen University, Zhuhai 519082, China}

\author{Z. P. Li}
\affiliation{School of Physical Science and Technology, Southwest University, Chongqing 400715, China}

\author{Peter Ring}
\affiliation{Department of Physik, Technische Universit\"{a}t M\"{u}nchen, D-85747 Garching, Germany}

\date{\today}

\begin{abstract}

A relativistic microscopic optical model potential for nucleon-nucleus scattering is developed based on the \emph{ab initio} relativistic Brueckner-Hartree-Fock (RBHF) theory with the improved local density approximation, which is abbreviated as the RBOM potential.
Both real and imaginary parts of the single-particle potentials in symmetric and asymmetric nuclear matter at various densities are determined uniquely in the full Dirac space.
The density distributions of the target nuclei are calculated by the covariant energy density functional theory with the density functional PC-PK1.
The central and spin-orbit terms of the optical potentials are quantitatively consistent with the relativistic phenomenological optical potentials.
The performance of the RBOM potential is evaluated by considering proton scattering with incident energy $E\leq 200$ MeV on five target nuclei, $\prescript{208}{}{\text{Pb}}$, $\prescript{120}{}{\text{Sn}}$, $\prescript{90}{}{\text{Zr}}$, $\prescript{48}{}{\text{Ca}}$, and $\prescript{40}{}{\text{Ca}}$.
Scattering observables including the elastic scattering angular distributions, analyzing powers, spin rotation functions, and reaction cross sections are analyzed.
Theoretical predictions show good agreements with the experimental data and the results derived from phenomenological optical potentials.
We anticipate that the RBOM potential can provide reference for other phenomenological and microscopic optical model potentials, as well as reliable descriptions for nucleon scattering on exotic nuclei in the era of rare-isotope beams.

\end{abstract}

\maketitle


\section{Introduction}\label{SecI}


Nuclear reaction constitutes an important field in nuclear physics, which is crucial not only for revealing nucleon-nucleon ($NN$) interactions and the structural and dynamic properties of nuclei~\cite{1985-Tanihata-PhysRevLett.55.2676}, but also for understanding the evolution of the stars and the origin of elements in the cosmos~\cite{1997-Wallerstein-RevModPhys.69.995, 2008-Schatz-Phys.Today, 2021-Cowan-RevModPhys.93.015002}.
In addition, nuclear reactions also have significant impacts on medical therapy, nuclear power, national security, etc.
Nucleon-nucleus scattering is one of the simplest and most important processes of nuclear reaction~\cite{2017-Sakaguchi-PPNP}.
For decades, numerous experimental data on scattering cross sections and polarization observables have been accumulated for different stable nuclei and a wide range of incident energies~\cite{[https://www-nds.iaea.org/exfor/]url-EXFOR}.


The optical model is an important theoretical tool for studying nucleon-nucleus scattering. It assumes that the complicated interaction between the incident nucleon and the target nucleus can be described with a complex mean field called the optical potential.
This divides the reaction flux into a part describing shape elastic scattering and a part covering all competing inelastic channels.
By solving the scattering equation of the incident nucleon with the optical potential, many experimental observables can be calculated, including the elastic scattering angular distributions, analyzing powers, spin rotation functions, reaction cross sections, etc.
Hence, the key to an optical model is constructing the optical potential.
For recent reviews on optical potentials, see Refs.~\cite{Dickhoff_2017-JPG, 2017-Sakaguchi-PPNP, 2019-Dickhoff-PPNP, Rotureau-2020-FP, 2022-JWHolt-arXiv, Hebborn_2023-JPG}.


The optical potentials can be constructed phenomenologically or microscopically.
The phenomenological methods within the non-relativistic framework~\cite{2009-Weppner-PhysRevC.80.034608} usually express the optical potential by a volume term, a surface term, and a spin-orbit term. 
Each term can be separated with an energy-dependent well depth and an energy-independent radial form factor.
The depth and geometry parameters are assumed to be functions of the incident energy as well as the mass number of the target nucleus.
They are determined using parameter adjustments to best fit the available experimental data.
The most widely used non-relativistic phenomenological optical potentials include the CH89~\cite{1991-Varner-Phys.Rep.} potential and the KD03~\cite{Koning-2003_Nucl.Phys.A713.231} potential.
Alternatively, the phenomenological methods within the relativistic framework are based on the Dirac phenomenology~\cite{1979-Arnold-Phys.Lett.B, 1979-Arnold-PhysRevC.19.917}, which starts from the Dirac equation for the single-particle motion of a nucleon in the nuclear medium.
Within the Dirac phenomenology, the real and imaginary parts of the scalar potential and vector potential are parametrized, from which the central and spin-orbit terms of the Schr\"odinger equivalent potential can be derived.
A distinct advantage of relativistic optical potentials is the natural inclusion of the spin-orbit potential, which arises from the difference between the contributions from the scalar and vector potentials.
The most widely used relativistic phenomenological optical potential currently is the global Dirac optical potential (GOP) constructed by Cooper, Hama, and Clark et al. in a series of papers~\cite{1987-Cooper-PhysRevC.36.2170, Hama-1990-PhysRevC.41.2737, Cooper-1993-PhysRevC.47.297, 2009-Cooper-PhysRevC.80.034605}.


Historically, the free parameters of phenomenological optical potentials have been fitted to elastic scattering data on stable targets. 
In recent years, with the advances of radioactive ion beam facilities worldwide, the nuclear landscape has been largely extended.
The accumulation of scattering data on exotic nuclei being produced requires a reliable theoretical optical model for analysis.
However, long-distance extrapolation and uncertainties are inevitable if phenomenological optical potentials are used to describe the scattering on these exotic nuclei.
For the description of the scattering phenomenology far away from stability, it is promising to develop microscopic optical model potentials based on realistic $NN$ interactions~\cite{Machleidt1989_ANP19-189, 2009-Epelbaum-RevModPhys.81.1773, 2011-Machleidt-Phys.Rep., 2017-Entem-PhysRevC.96.024004, 2022-LuJX-PhysRevLett.128.142002}, whose parameters are nicely calibrated with respect to deuteron properties and $NN$ scattering data in the free space.


In principle, constructing a microscopic optical potential requires the full solution of the $A+1$ nuclear quantum many-body problem with $A$ the mass number of the target nucleus.
This is quite complicated and beyond present capabilities.
In practice, several methods based on suitable approximations have been developed.
Considerable progress in this field has been achieved in Refs.~\cite{Deb2001-PhysRevLett.86.3248, Quaglioni2008-PhysRevLett.101.092501, Hagen2012-PhysRevC.86.021602, Vorabbi-2016-PhysRevC.93.034619, Lynn2016-PhysRevLett.116.062501, Rotureau-2017-PhysRevC.95.024315, 2019-Idini-PhysRevLett.123.092501, Whitehead-2021-PhysRevLett.127.182502}.
Among these methods, the folding method expresses the optical potential as the $NN$ scattering amplitudes folded with the nuclear density distributions of the target.
This method has been applied with $T$ matrix~\cite{Li-2008-PhysRevC.78.014603, Vorabbi-2016-PhysRevC.93.034619, KuangY-2023-EPJA} and $G$ matrix~\cite{2008-Furumoto-PhysRevC.78.044610, Furumoto-2019-PhysRevC.99.034605}.
The local density approximation (LDA) is another effective method to construct microscopic optical potentials, in which the basic idea is that the optical potentials are equivalent to the single-particle potentials in nuclear matter~\cite{1959-Bell-PhysRevLett.3.96}.
The LDA requires a self-consistent solution for symmetric and asymmetric nuclear matter a priori.
In practice, through the improved version, LDA has been combined with \textit{ab initio} calculations based on the non-relativistic Brueckner-Hartree-Fock (BHF) theory~\cite{Bauge-1998-PhysRevC.58.1118, Bauge-2001-PhysRevC.63.024607}, the many-body perturbation theory (MBPT)~\cite{Whitehead-2019-PhysRevC.100.014601,Whitehead-2021-PhysRevLett.127.182502}, and the relativistic Brueckner-Hartree-Fock (RBHF) theory~\cite{1994-GQLi-Nucl.Phys.A, 2012-RuiRuiXU-PhysRevC.85.034613, XuRR-2016-PRC-94-034606}.
Especially, the combination of the LDA and RBHF theory, supplemented with minor parameter adjustments, has yielded a successful relativistic microscopic optical potential called CTOM~\cite{XuRR-2016-PRC-94-034606, 2017-RuiRuiXu-EPJWC}, with which good descriptions of the experimental data for nucleon-nucleus scattering over a broad range of targets and a large region of energies, have been obtained.


For the relativistic microscopic optical potentials constructed with the LDA, it is crucial to accurately determine the scalar and vector components of single-particle potentials in nuclear matter.
This can only be accomplished in the full Dirac space, where the positive-energy states (PESs) and negative-energy states (NESs) are considered simultaneously~\cite{1989-Nuppenau-Nucl.Phys.A, VanGiai2010JPG}.
In previous studies including CTOM, the RBHF theory for infinite nuclear matter has been solved with the momentum-independence approximation~\cite{Brockmann1990_PRC42-1965} or the projection method~\cite{1999-Gross-Boelting-Nucl.Phys.A}, where the effects from NESs are neglected.
This leads to ambiguities of the single-particle potentials and, hence, the constructed optical potentials.


Recently, we achieved the self-consistent RBHF solutions for symmetric and asymmetric nuclear matter in the full Dirac space~\cite{WANG-SB2021_PRC103-054319, Tong_2022-AstrophysicsJ930.137, WANG-SB2022_PhysRevC.105.054309}.
The scalar and vector components of the single-particle potential are obtained without ambiguities, which solves the long-standing problem of being unable to determine the single-particle potentials uniquely in RBHF calculations.
The RBHF theory in the full Dirac space has been successfully applied to study the effective masses in nuclear matter~\cite{2022-Wang-SIBO-PhysRevC.106.L021305, 2023-WangSB-PhysRevC.108.L031303}, the properties of $^{208}$Pb with a liquid droplet model~\cite{2023-Tong-PhysRevC.107.034302}, and neutron star properties~\cite{Tong_2022-AstrophysicsJ930.137, 2022-Wang-PhysRevC.106.045804, 2023-QuXY-SciChina}.


This work constructs a relativistic microscopic optical model potential by combining the RBHF theory in the full Dirac space with the LDA.
The elastic scattering angular distributions, analyzing powers, spin rotation functions, and reaction cross sections for proton-nucleus scattering will be investigated to evaluate the performance of the optical potential.
The description of the neutron-nucleus scattering will be studied in a forthcoming paper.

This paper is organized as follows.
The theoretical formalism of the RBHF theory in the full Dirac space, the construction of microscopic optical potential RBOM, and the extraction of scattering observables are briefly introduced in Sec.~\ref{sec:theo}.
In Sec.~\ref{sec:exp.}, we present the single-particle potentials in nuclear matter, the optical potentials for proton-nucleus scattering, the scattering observables, and the uncertainty quantification for RBOM.
Finally in Sec.~\ref{sec:summary}, the overall discussion is summarized and some prospects are given.

\section{Theoretical framework} \label{sec:theo}

\subsection{The relativistic Brueckner-Hartree-Fock theory in the full Dirac space}\label{sec:RBHF}

Within the RBHF theory, the single-particle motion of a nucleon in infinite nuclear matter with the rest mass $M$, momentum $\bm{p}$, and energy $E_{\bm{p},\tau}$ is described with the Dirac equation
\begin{equation}\label{eq:DiracEquation}
  \left[ \bm{\alpha}\cdot\bm{p}+\beta \left(M+\mathcal{U}_\tau\right) \right] u_\tau(\bm{p},s) 
  = E_{\bm{p},\tau}u_{\tau}(\bm{p},s),
\end{equation}
where $\bm{\alpha}$ and $\beta$ are the Dirac matrices, $s$ is the spin, and $\tau$ denotes the neutron $(n)$ or the proton $(p)$.
Due to the translational invariance and the parity conservation in homogeneous nuclear matter, the Lorentz structure of the single-particle potential operator $\mathcal{U}_\tau$ can be decomposed as~\cite{1986-Serot-ANP}
\begin{equation}\label{eq:SPP}
  \mathcal{U}_\tau(\bm{p}) = U_{S,\tau}(p)+ \gamma^0U_{0,\tau}(p) + \bm{\gamma\cdot\hat{p}}U_{V,\tau}(p).
\end{equation}
Here $U_{S,\tau}$ is the scalar potential, $U_{0,\tau}$ and $U_{V,\tau}$ are the timelike and spacelike components of the vector potential, and $\hat{\bm{p}}=\bm{p}/|\bm{p}|=\bm{p}/p$ is the unit vector parallel to the momentum $\bm{p}$.

By introducing the effective mass $M^*_{\bm{p},\tau}=M+U_{S,\tau}(p)$, effective momentum $\bm{p}^*_\tau=\bm{p}+\hat{\bm{p}}U_{V,\tau}(p)$, and effective energy $E^*_{\bm{p},\tau}=E_{\bm{p},\tau}-U_{0,\tau}(p)$, one can rewrite the in-medium Dirac equation \eqref{eq:DiracEquation} in the form of the free Dirac equation.
This allows to obtain the PES and the NES analytically
\begin{subequations}\label{eq:DiracSpinor}
  \begin{align}
    u_\tau(\bm{p},s) =&\ \sqrt{\frac{E_{\bm{p},\tau}^*+M_{\bm{p},\tau}^*}{2M_{\bm{p},\tau}^*}}
      \bbm 1 \\ \frac{\bm{\sigma}\cdot\bm{p}^*_\tau}{E_{\bm{p},\tau}^*+M_{\bm{p},\tau}^*}\ebm \chi_s\chi_\tau, 
      \label{eq:DiracSpinor-u} \\
    v_\tau(\bm{p},s) =&\ \gamma^5u_\tau(\bm{p},s),
      \label{eq:DiracSpinor-v}
  \end{align}
\end{subequations}
where $\chi_s$ and $\chi_\tau$ are the spin and isospin wavefunction of a nucleon.

In the full Dirac space, the matrix elements of the single-particle potential operator can be expressed by $U_S$, $U_0$, and $U_V$~\cite{1981-Anastasio-PRC, 1988-Poschenrieder-PRC}:
\begin{subequations}\label{eq:Sigma++-+--}
  \begin{align}
    \Sigma^{++}_\tau(p)
    =&\   \bar{u}_\tau(\bm{p},1/2) \mathcal{U}_\tau(\bm{p}) u_\tau(\bm{p},1/2) 
          = U_{S,\tau}(p) + \frac{E^*_{\bm{p},\tau}}{M^*_{\bm{p},\tau}} U_{0,\tau}(p) 
                    + \frac{p^*_\tau}{M^*_{\bm{p},\tau}} U_{V,\tau}(p), \label{eq:Sigma++def}\\
    \Sigma^{-+}_\tau(p)
    =&\  \bar{v}_\tau(\bm{p},1/2) \mathcal{U}_\tau(\bm{p}) u_\tau(\bm{p},1/2) 
          = \frac{p^*_\tau}{M^*_{\bm{p},\tau}} U_{0,\tau}(p)
            + \frac{E^*_{\bm{p},\tau}}{M^*_{\bm{p},\tau}} U_{V,\tau}(p), \label{eq:Sigma-+def}\\
    \Sigma^{--}_\tau(p)
    =&\   \bar{v}_\tau(\bm{p},1/2) \mathcal{U}_\tau(\bm{p}) v_\tau(\bm{p},1/2)
          = - U_{S,\tau}(p) + \frac{E^*_{\bm{p},\tau}}{M^*_{\bm{p},\tau}} U_{0,\tau}(p)
                    + \frac{p^*_\tau}{M^*_{\bm{p},\tau}} U_{V,\tau}(p). \label{eq:Sigma--def}
  \end{align}
\end{subequations}
From these matrix elements one can derive different components of the single-particle potential operator
\begin{subequations}\label{eq:Sigma2US0V}
  \begin{align}
    U_{S,\tau}(p) = &\ \frac{\Sigma^{++}_\tau(p)-\Sigma^{--}_\tau(p)}{2},\label{eq:Sigma2Us}\\
    U_{0,\tau}(p) = &\ \frac{E^*_{\bm{p},\tau}}{M^*_{\bm{p},\tau}}\frac{\Sigma^{++}_\tau(p)+\Sigma^{--}_\tau(p)}{2}
    					 - \frac{p^*_\tau}{M^*_{\bm{p},\tau}}\Sigma^{-+}_\tau(p),\label{eq:Sigma2U0}\\
    U_{V,\tau}(p) = &\ -\frac{p^*_\tau}{M^*_{\bm{p},\tau}}\frac{\Sigma^{++}_\tau(p)+\Sigma^{--}_\tau(p)}{2} 
    				  + \frac{E^{*}_{\bm{p},\tau}}{M^*_{\bm{p},\tau}} \Sigma^{-+}_\tau(p).\label{eq:Sigma2UV}
  \end{align}
\end{subequations}
In parallel, the matrix elements in Eq.~\eqref{eq:Sigma++-+--} can be calculated with the integral of the effective interaction between nucleons, i.e., the $G$ matrix in the RBHF theory.
As the infinite summation of the bare $NN$ interaction $V$, the $G$ matrix is obtained by solving the Thompson equation~\cite{Brockmann1990_PRC42-1965} in nuclear medium
\begin{equation}\label{eq:ThomEqu}
  \begin{split}
  G_{\tau\tau'}(\bm{q}',&\bm{q}|\bm{P},W)
  = V_{\tau\tau'}(\bm{q}',\bm{q}|\bm{P})
  + \int \frac{d^3k}{(2\pi)^3}
  V_{\tau\tau'}(\bm{q}',\bm{k}|\bm{P}) \\
    & \times
    \frac{Q_{\tau\tau'}(\bm{k},\bm{P})}{W-E_{\bm{P}+\bm{k},\tau}-E_{\bm{P}-\bm{k},\tau'}} 
     G_{\tau\tau'}(\bm{k},\bm{q}|\bm{P},W),
  \end{split}
\end{equation}
Here $\bm{P}=\frac{1}{2}({\bm k}_1+{\bm k}_2)$ is the half of the total momentum, and $\bm{k}=\frac{1}{2}({\bm k}_1-{\bm k}_2)$ is the relative momentum of the two interacting nucleons with momenta ${\bm k}_1$ and ${\bm k}_2$.
The initial, intermediate, and final relative momenta of the two nucleons scattering in nuclear matter are denoted as $\bm{q}, \bm{k}$, and $\bm{q}'$, respectively.
The starting energy $W$ denotes the sum of the single-particle energies of two nucleons in the initial states~\cite{WANG-SB2021_PRC103-054319}.
The Pauli operator $Q_{\tau\tau'}(\bm{k},\bm{P})$ prohibits the nucleons from scattering to the occupied states.
For the sake of brevity, the indexes for PESs or NESs in Eq.~\eqref{eq:ThomEqu} are suppressed.

In practical calculations, the Thompson equation \eqref{eq:ThomEqu} is decomposed into partial waves with the total angular momentum $J$ in the helicity scheme and reduced to a one-dimensional integral equation over the relative momentum $k$.
To avoid the mixture of different $J$-channels, angular averaging has to be performed on the angles between vectors $\bm{k}$ and $\bm{P}$ before partial wave decomposition~\cite{Brockmann1990_PRC42-1965, 2003-Alonso-PhysRevC.67.054301}.
Specifically, the Pauli operator $Q_{\tau\tau'}(\bm{k},\bm{P})$ is replaced with $Q^{\text{av}}_{\tau\tau'}(k,P)$, and the energy denominator $W-E_{\bm{P}+\bm{k},\tau}-E_{\bm{P}-\bm{k},\tau'}$ is replaced with $W-E^{\text{av}}_{\tau}(k,P)-E^{\text{av}}_{\tau'}(k,P)$.

Constructing a microscopic optical potential requires a unified treatment of the single-particle states below and above the Fermi surface, i.e., the \textit{continuous choice}~\cite{Jeukenne1976_PR25-83, Baldo-2007-JPG.34.R243}.
This would bring a singularity in Eq.~\eqref{eq:ThomEqu} whenever the starting energy $W$ is larger than twice the Fermi energy.
Adding an infinitesimal $i\epsilon$ in the denominator, Eq.~\eqref{eq:ThomEqu} can be rewritten as a complex equation~\cite{WANG-SB2022_PhysRevC.105.054309}
\begin{equation}\label{eq:ThomEqu-final}
  \begin{split}
    &\la q'|G^J_{\tau\tau'}(P,W)|q\ra \\
    = \la q'&|V^J_{\tau\tau'}(P)|q\ra 
    + \int \frac{k^2dk}{(2\pi)^3} \la q'|V^J_{\tau\tau'}(P)|k\ra 
      \frac{Q^{\mathrm{av}}_{\tau\tau'}(k,P)}{W - E^{\text{av}}_\tau(k,P) - E^{\text{av}}_{\tau'}(k,P)}  \la k|G^J_{\tau\tau'}(P,W)|q\ra \\
      - &\ \la q'|V^J_{\tau\tau'}(P)|k_0\ra  Q^{\mathrm{av}}_{\tau\tau'}(k_0,P) 
        \frac{W + E^{\text{av}}_\tau(k_0,P)+E^{\text{av}}_{\tau'}(k_0,P)}{A_{\tau\tau'}(k_0,P)}\\
      &\ \times \left[ k^2_0 \int \frac{dk}{(2\pi)^3} \frac{1}{4(k^2_0-k^2)} + \frac{i\pi}{(2\pi)^3} \frac{k_0}{8} \right] 
        \la k_0|G^J_{\tau\tau'}(P,W)|q\ra .
  \end{split}
\end{equation}
Here, the quantity $k_0$ denotes the position of the singularity for the energy denominator.
The value $A_{\tau\tau'}(k_0,P)=\lim_{k\rightarrow k_0} \frac{W^2-\left[E^{\text{av}}_\tau(k,P)+E^{\text{av}}_{\tau'}(k,P)\right]^2}{4(k^2_0-k^2)} $ can be calculated by L'Hospital's rule.

Once the $G$ matrix are prepared, the matrix elements of the single-particle potential operator can be obtained as 
\begin{subequations}\label{eq:Gm2Sigma}
  \begin{align}
    \Sigma^{++}_\tau(p) = &\ \sum_{s'\tau'} \int^{k^{\tau'}_F}_0 \frac{d^3p'}{(2\pi)^3}
          \frac{M^*_{\bm{p}',\tau'}}{E^*_{\bm{p}',\tau'}}
          \la \bar{u}_\tau(\bm{p},1/2) \bar{u}_{\tau'}(\bm{p}',s')| 
          \bar{G}^{++++}(W)| u_\tau(\bm{p},1/2)u_{\tau'}(\bm{p}',s')\ra, \label{eq:Sigma++} \\ 
    \Sigma^{-+}_\tau(p) = &\ \sum_{s'\tau'} \int^{k^{\tau'}_F}_0 \frac{d^3p'}{(2\pi)^3}
          \frac{M^*_{\bm{p}',\tau'}}{E^*_{\bm{p}',\tau'}}
          \la \bar{v}_\tau(\bm{p},1/2) \bar{u}_{\tau'}(\bm{p}',s')| 
          \bar{G}^{-+++}(W)| u_\tau(\bm{p},1/2)u_{\tau'}(\bm{p}',s')\ra, \label{eq:Sigma-+} \\ 
    \Sigma^{--}_\tau(p) = &\ \sum_{s'\tau'} \int^{k^{\tau'}_F}_0 \frac{d^3p'}{(2\pi)^3}
          \frac{M^*_{\bm{p}',\tau'}}{E^*_{\bm{p}',\tau'}}
          \la \bar{v}_\tau(\bm{p},1/2) \bar{u}_{\tau'}(\bm{p}',s')| 
          \bar{G}^{-+-+}(W)| v_\tau(\bm{p},1/2)u_{\tau'}(\bm{p}',s')\ra. \label{eq:Sigma--}
  \end{align}
\end{subequations}
Here $k^{\tau}_F=\left(3\pi^2\rho/2\right)^{1/3}(1+\tau_3\alpha)^{1/3}$ is the Fermi momentum, where $\rho=\rho_n+\rho_p$ and $\alpha = (\rho_n-\rho_p)/\rho$ are the total density and isospin asymmetry parameter.
For the neutron (proton), $\tau_3 = +1(-1)$.
In Eq.~\eqref{eq:Gm2Sigma}, $\bar{G}$ is the antisymmetric $G$ matrix, where the superscripts denote the PESs or NESs.

The extraction of different components of single-particle potentials depends on the $G$ matrix, and the calculation of $G$ matrix, in turn, depends on the single-particle potentials.
Therefore, Eqs.~\eqref{eq:DiracEquation}, \eqref{eq:Sigma2US0V}, \eqref{eq:ThomEqu-final}, and \eqref{eq:Gm2Sigma} constitute a coupled system which needs a self-consistent solution.
We emphasize that the solution of Eq.~\eqref{eq:ThomEqu-final} yields a complex $G$ matrix, from which the matrix elements of the single-particle potential operator \eqref{eq:Gm2Sigma} and the components of single-particle potentials \eqref{eq:Sigma2US0V} are also complex.
Details on the RBHF theory in the full Dirac space can be found in Refs.~\cite{WANG-SB2021_PRC103-054319, Tong_2022-AstrophysicsJ930.137, WANG-SB2022_PhysRevC.105.054309}.

For simplicity, the spacelike part of the vector potential $U_V$ in Eq.~\eqref{eq:DiracEquation} is absorbed by redefining the scalar and the timelike parts of the vector potential~\cite{2012-RuiRuiXU-PhysRevC.85.034613}
\begin{equation}\label{eq:Uvbar}
  \tilde{U}_{S,\tau} = \frac{U_{S,\tau}-\bar{U}_{V,\tau} M}{1+\bar{U}_{V,\tau}},\qquad 
  \tilde{U}_{0,\tau} = \frac{U_{0,\tau}+\bar{U}_{V,\tau}E_{\bm{p},\tau}}{1+\bar{U}_{V,\tau}},
\end{equation}
where $\bar{U}_{V,\tau}(p)=U_{V,\tau}(p)/p$ is a dimensionless quantity.
Discussions of optical potentials in the following only involve redefined quantities and we continue using the notation $U_{S,\tau}$ and $U_{0,\tau}$ instead of $\tilde{U}_{S,\tau}$ and $\tilde{U}_{0,\tau}$.

\subsection{Microscopic optical potential with the improved local density approximation}

In the relativistic scheme, the Dirac equation for the single-particle motion of a projectile in the field generated by the target can be written as 
\begin{equation}\label{eq:Dirac-sca}
  \left[ \bm{\alpha}\cdot\bm{p} + \beta(M+U_S) \right] \psi(\bm{r}) = (E - U_0) \psi(\bm{r}).
\end{equation}
Here $E=\sqrt{p^2+M^2}$ is the energy of the projectile in the center-of-mass (c.m.) frame of the nucleon and nucleus.
$p\equiv|\bm{p}|$ stands for the momentum of the projectile in the c.m. frame, which is related to the target mass $M_t$ and the incident energy $E_{\text{lab}}$ with the relativistic kinematics through~\cite{2021-Arellano-CPC}
\begin{equation}
	p = \frac{M_t^2\left(E^2_{\text{lab}}+2ME_{\text{lab}}\right)}{2M_tE_{\text{lab}} + (M+M_t)^2}.
\end{equation}

The scattering-state Dirac equation for the four-component spinor $\psi=[\varphi, \chi]^T$ is equivalent to two coupled equations for the upper two-component spinor $\varphi$ and the lower one $\chi$.
Usually, a Schr\"odinger equivalent equation for the upper component is derived by eliminating the lower component in Eq.~\eqref{eq:Dirac-sca} in a standard way~\cite{1981-Arnold-PhysRevC.23.1949}
\begin{equation}\label{eq:phi}
  \left[ - \frac{\nabla^2}{2E} + V_{\text{cent}} (r) + V_{\text{so}} (r) \bm{\sigma}\cdot\bm{l} 
  + V_{\text{Darwin}}(r)\right]  \phi(\bm{r}) = \frac{E^2-M^2}{2E} \phi(\bm{r}).
\end{equation}
Here the wave function $\phi$ is related to the upper component $\varphi$ by $\phi = \left[ D/(E+M)\right]^{-1/2}\varphi$  with $D=M+E+U_S-U_0$. 
In Eq.~\eqref{eq:phi}, the central term $V_{\text{cent}}$, spin-orbit term $V_{\text{so}}$, and Darwin term $V_{\text{Darwin}}$ are related to the scalar and vector potentials through
\begin{subequations}
  \begin{align} 
    V_{\text{cent}} = &\  \frac{M}{E} U_S + U_0 + \frac{1}{2E}(U^2_S - U^2_0), \label{eq:cent}\\
    V_{\text{so}} = &\ -\frac{1}{2ErD} \frac{dD}{dr},  \label{eq:vso}\\
    V_{\text{Darwin}} =&\  \frac{3}{8ED^2} \left(\frac{dD}{dr}\right)^2 - \frac{1}{2ErD} \frac{dD}{dr}
  - \frac{1}{4ED} \frac{d^2D}{dr^2}.
  \end{align}
\end{subequations}
It can be seen that, the central term $V_{\text{cent}}$ is determined by the cancellation of the scalar and vector potentials, while the spin-orbit term $V_{\text{so}}$ depends mainly on the derivative of the difference between the scalar and vector potentials. 
Since $U_S$ and $U_0$ have opposite signs, this automatically gives a large spin-orbit potential.
For the proton-nucleus scattering, the Coulomb potential $V_{\text{coul}}$ is added on the central term $V_{\text{cent}}$~\cite{XuRR-2016-PRC-94-034606}.
This allows our optical potential being used conveniently in non-relativistic nucleon-nucleus scattering codes.
Although the nucleon-nucleus scattering is described with the Schr\"odinger equivalent equation \eqref{eq:phi}, the relativity is important in the rest of our investigations because of the large scalar and vector fields and the automatic consideration of the spin-orbit interactions etc.

In this work, the Schr\"odinger equivalent optical potentials of a nucleon scattering off a target nucleus are obtained by means of the LDA, where the optical potential at radial distance $r$ can be directly related to the single-particle potentials in nuclear matter with density $\rho$ and isospin asymmetry parameter $\alpha$ locally
\begin{equation}\label{eq:LDA}
	U_{\text{LDA}}(r,\varepsilon) = U_{\text{NM}}(\varepsilon,\rho(r),\alpha(r)).
\end{equation}
Here the quantity $U_{\text{NM}}$ can be the scalar potential, vector potential, or their combinations.
In the right-hand side of Eq.~\eqref{eq:LDA}, $\varepsilon\equiv\varepsilon_{\bm{p},\tau}=E_{\bm{p},\tau}-M$ is the single-particle energy subtracting the rest mass of a nucleon in nuclear matter.
In the left-hand side, $\varepsilon=\sqrt{p^2+M^2} - M $ is the kinetic energy of the projectile in the c.m. frame.
In this work, the radial profiles of the nucleon density $\rho(r)$ and the isospin asymmetry $\alpha(r)$ are obtained by the covariant density functional theory with the density functional PC-PK1~\cite{2010-ZhaoPW-PhysRevC.82.054319}.
The Coulomb potential $V_{\text{coul}}$ is also calculated with the charge density distribution obtained with PC-PK1.

To consider the finite range correction of the $NN$ interaction, a Gaussian form factor is introduced to improve the optical potentials, i.e., the so-called improved local density approximation (ILDA) is adopted~\cite{1977-Jeukenne-PhysRevC.16.80, Bauge-1998-PhysRevC.58.1118}.
Finally, with the range parameter $t$ representing the effective range of the $NN$ interaction, the optical potential is obtained by the following integral
\begin{equation}\label{eq:ILDA}
	U_{\text{ILDA}}(r,\varepsilon) = (t\sqrt{\pi})^{-3} \int U_{\text{LDA}}(r',\varepsilon) 
		\exp\left( -|\bm{r}-\bm{r}'|^2/t^2\right) d^3 r'.
\end{equation}
We call the optical potential $U_{\text{ILDA}}$ in this work as the RBOM potential.
It should be emphasized that $t$ is the only adjustable parameter in this optical potential.
In previous works, this parameter is usually treated as a free parameter and fitted to experimental data.
In Ref.~\cite{Whitehead-2021-PhysRevLett.127.182502}, $t_{\text{c}}=1.22$\ fm for the central potentials and $t_{\text{so}}=0.98$\ fm for the spin-orbit potentials are obtained.
Similarly, in Ref.~\cite{XuRR-2016-PRC-94-034606}, the effective range factor $t$ in Eq.~\eqref{eq:ILDA} is determined as 1.35 fm for proton-nucleus scatting and 1.45 fm for neutron-nucleus scattering.
In the following discussions, $t=1.3$\ fm is adopted, and the uncertainties of the scattering observables from the varies of the range parameter $t$ will be discussed afterwards.

\subsection{The extraction of scattering observables}

A partial-wave expansion of Eq.~\eqref{eq:phi} leads to the radial equation
\begin{equation}\label{eq:upm}
  \left\{ \frac{d^2}{dr^2} + p^2 - 2E\left[ V_{\text{cent}}(r) + V_{\text{Darwin}}(r) + l_\pm V_{\text{so}} (r) \right]
  - \frac{l(l+1)}{r^2} \right\} u^{\pm}_l(pr) = 0.
\end{equation}
Here $l_+ = l$ and $l_-=-l-1$. This radial equation can be solved with the standard Numerov method.
At large enough $r$, the reduced radial functions $u^{\pm}_l$ is matched to the appropriate linear combinations of asymptotic functions via 
\begin{equation}
   u^{\pm}_{l}(pr) \sim F_l(pr) + \mathcal{C}^{\pm}_{l} [G_l(pr)+iF_l(pr)],
\end{equation}
where coefficients $\mathcal{C}^{\pm}_{l}$ are to be determined.
The $F_l$ and $G_l$ are Coulomb functions in a problem without nuclear potentials~\cite{1974-Barnett-CPC}.

Considering collisions of spin-$1/2$ nucleons with spin-0 target, the scattering amplitudes $A(\theta)$ and $B(\theta)$ with the scattering angle in the c.m. frame $\theta\equiv\theta_{\text{c.m.}}$ are obtained from the relations~\cite{2021-Arellano-CPC}
\begin{subequations}
  \begin{align}
    A(\theta) =&\ f_c(\theta) + \frac{1}{p} \sum_l e^{2i\sigma_l} [(l+1)\mathcal{C}^+_l + l\mathcal{C}^-_l] P_l(\cos\theta),\\
    B(\theta) =&\ \frac{i}{p} \sum_{l}e^{2i\sigma_l} [\mathcal{C}^+_l - \mathcal{C}^-_l] P^1_l(\cos\theta).
  \end{align}
\end{subequations}
Here $P_l$ and $P^1_l$ denote the Legendre polynomials and their derivatives.
The Coulomb amplitude $f_c(\theta)$ is given by
\begin{equation}
  f_c(\theta) = \frac{-\eta}{2p\sin^2(\theta/2)}\exp\{ -i\eta \ln[\sin^2(\theta/2)]+2i\sigma_0\},
\end{equation}
where $\sigma_l=\mathrm{arg}\Gamma(l+1+i\eta)$ are the Coulomb phase shifts with $\eta = (\mu Ze^2)/p$. 
Here $\mu$ denotes the nucleon-nucleus reduced mass and the quantity $Z$ is the charge of the target.

From the scattering amplitudes, it is straightforward to deduce the experimental observables, including the differential cross section $d\sigma/d\Omega(\theta)$, analyzing power $A_y(\theta)$, and spin rotation function $Q(\theta)$~\cite{1991-Horowitz-ComNuclPhys}:
\begin{subequations}
  \begin{align}
    \frac{d\sigma}{d\Omega}(\theta)=&\ |A(\theta)|^2 + |B(\theta)|^2,\\
    A_y(\theta) =&\ \frac{2\mathrm{Re}[A^*(\theta)B(\theta)]}{d\sigma/d\Omega},\\
    Q  (\theta) =&\ \frac{2\mathrm{Im}[A^*(\theta)B(\theta)]}{d\sigma/d\Omega}.
  \end{align}
\end{subequations}
The elastic cross section $\sigma_{\text{el}}$, reaction cross section $\sigma_{\text{reac}}$, and total cross section $\sigma_{\text{tot}}$ are evaluated with~\cite{2020-Blanchon-CPC}
\begin{subequations}
  \begin{align}
    \sigma_{\text{el}}   =&\ \frac{\pi}{k^2} \sum_{l} \left[ (l+1) |1-S^+_l|^2 + l |1-S^-_l|^2 \right], \\
	  \sigma_{\text{reac}} =&\ \frac{\pi}{k^2} \sum_{l} \left[ (l+1) \left( 1-|S^+_l|^2 \right) + l \left( 1-|S^-_l|^2 \right) \right], \\
    \sigma_{\text{tot}}  =&\ \frac{\pi}{k^2} \sum_{l} \left[ (l+1) \left(1-\text{Re}S^+_l\right) + l \left(1-\text{Re}S^-_l\right) \right], 
  \end{align}
\end{subequations}
where the $S$-matrices $S^\pm_l$ are related to the coefficients $\mathcal{C}^{\pm}_{l}$ by $S^\pm_l = 1 + 2i \mathcal{C}^\pm_l$.

\section{Results and discussion}\label{sec:exp.}

\subsection{Single-particle potentials in nuclear matter}


\begin{figure}[htbp]
  \centering
  \includegraphics[width=16.0cm]{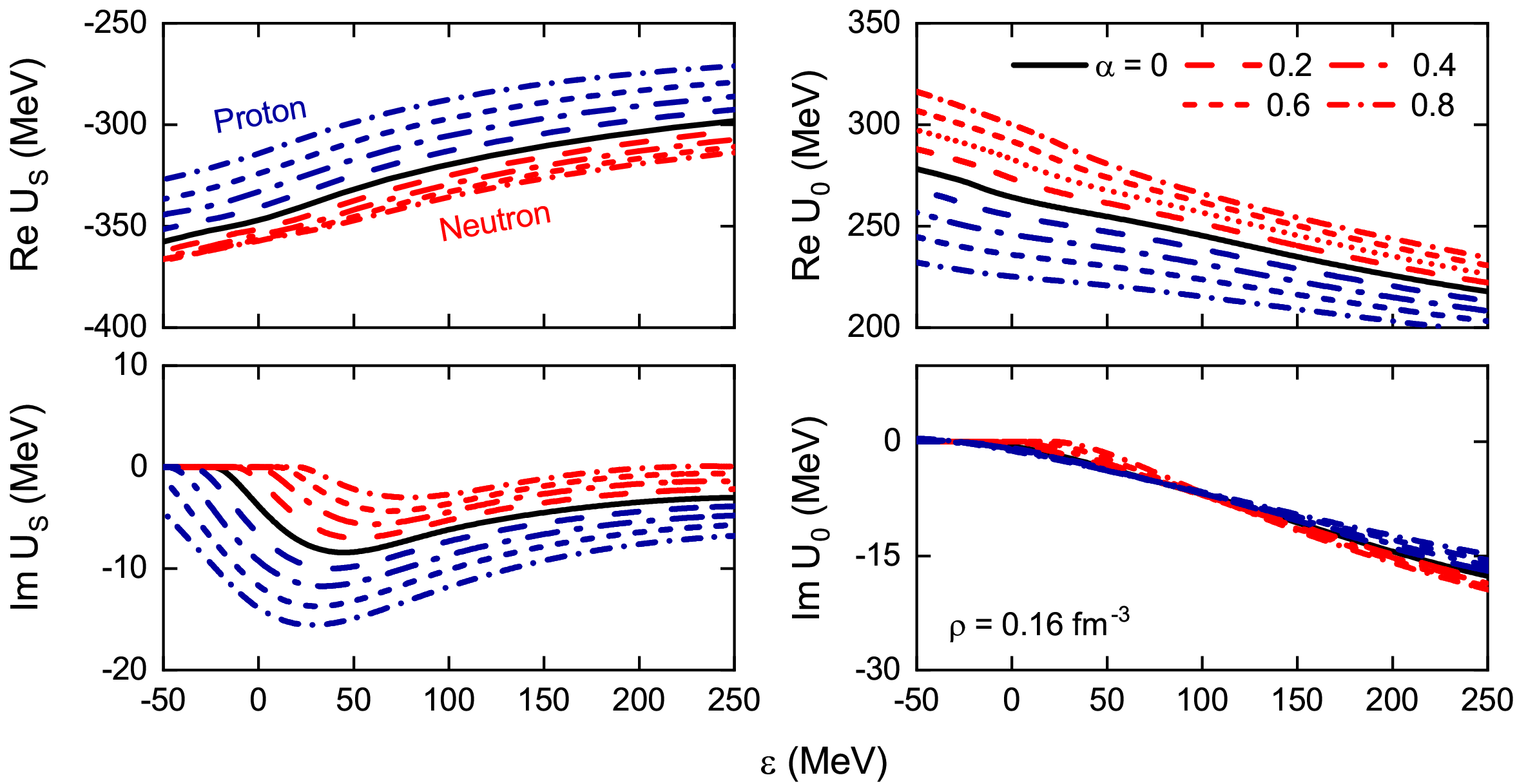}
  \caption{(Color online) The real and imaginary parts of the scalar potential $U_{S,\tau}$ and vector potential $U_{0,\tau}$ as functions of single-particle energy $\varepsilon$ with the asymmetry parameter $\alpha$ ranging from 0 to 0.8. The density is chosen at $\rho=0.16\ \text{fm}^{-3}$.}
  \label{label-Fig1}
\end{figure}

In this subsection, we focus on the single-particle potentials of a nucleon in nuclear matter calculated by the RBHF theory in the full Dirac space with the realistic Bonn A potential~\cite{Brockmann1990_PRC42-1965}.
The binding	energies per particle and saturation properties for nuclear matter in a wide range of densities and isospin asymmetry parameters can be found in Refs.~\cite{WANG-SB2021_PRC103-054319,Tong_2022-AstrophysicsJ930.137}.
In Fig.~\ref{label-Fig1}, we show the real and imaginary parts of the scalar potential $U_{S,\tau}$ and vector potential $U_{0,\tau}$ as functions of single-particle energy $\varepsilon$ with the asymmetry parameter $\alpha$ ranging from 0 to 0.8. 
The density is chosen at the empirical saturation density $\rho=0.16\ \text{fm}^{-3}$.
The real parts are of the order of several hundreds of MeV, while the imaginary parts are one order smaller and vanish for single-particle states below the Fermi surface.
The isospin dependence for Im$U_{0,\tau}$ is found to be much weaker than the others, which is related to the fact that the isospin dependences of Im$U_{0,\tau}$ and Im$\bar{U}_{V,\tau}$ in Eq.~\eqref{eq:Uvbar} are largely and accidentally canceled.
With the increase of single-particle energy, the magnitudes of both Re$U_{S,\tau}$ and Re$U_{0,\tau}$ decrease, while that of Im$U_{0,\tau}$ increases. For Im$U_{S,\tau}$, a nonmonotonical trend is found, especially for the proton with large isospin asymmetry.
Similarly to what has been observed in the precursor studies for CTOM in Ref.~\cite{2012-RuiRuiXU-PhysRevC.85.034613}, the imaginary part of the scalar potential $\text{Im}U_{S,\tau}$ obtained in this work is negative, especially for protons, which is different from the positive results usually found in phenomenological Dirac optical potentials~\cite{1987-Cooper-PhysRevC.36.2170, Hama-1990-PhysRevC.41.2737, Cooper-1993-PhysRevC.47.297}.


\begin{figure}[htbp]
  \centering
  \includegraphics[width=16.0cm]{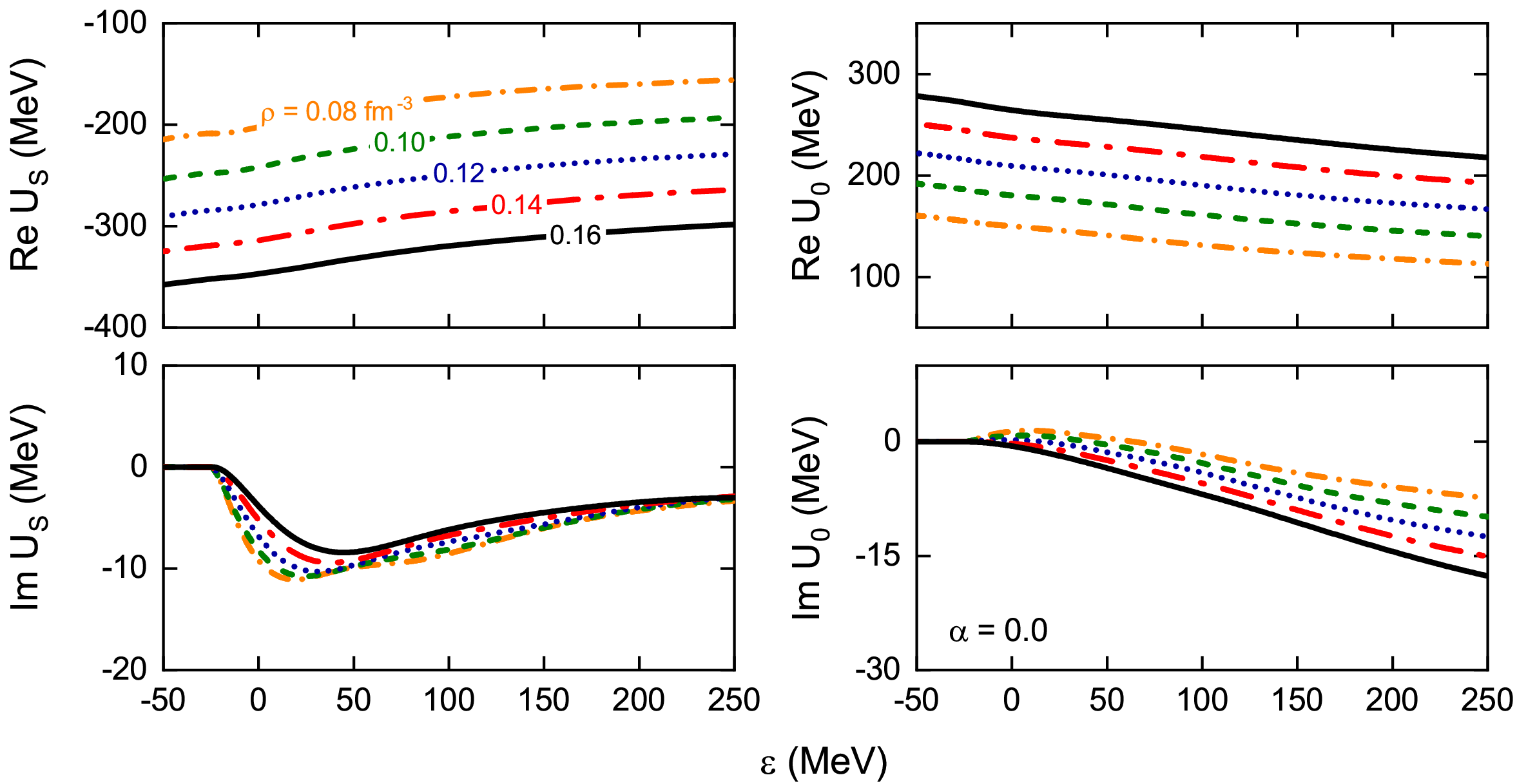}
  \caption{(Color online) The real and imaginary parts of the scalar potential $U_{S,\tau}$ and vector potential $U_{0,\tau}$ as functions of single-particle energy $\varepsilon$ with the density ranging from 0.08 to 0.16 fm$^{-3}$. The asymmetry parameter is chosen at $\alpha=0$.}
  \label{label-Fig2}
\end{figure}

In Fig.~\ref{label-Fig2}, we show the energy dependence for the real and imaginary parts of the scalar potential $U_{S,\tau}$ and vector potential $U_{0,\tau}$ with the density ranging from 0.08 to 0.16\ $\text{fm}^{-3}$. 
The isospin asymmetry parameter is chosen at $\alpha=0$.
With the decrease of the density, the amplitudes of the real parts of the scalar and vector potentials decrease towards the vanishing physical limits at extreme low densities.
With the increase of the single-particle energy, the difference in $\text{Im}U_0$ among different densities is also increasing, while the difference in $\text{Im}U_S$ is decreasing overall.


\begin{figure}[htbp]
  \centering
  \includegraphics[width=16.0cm]{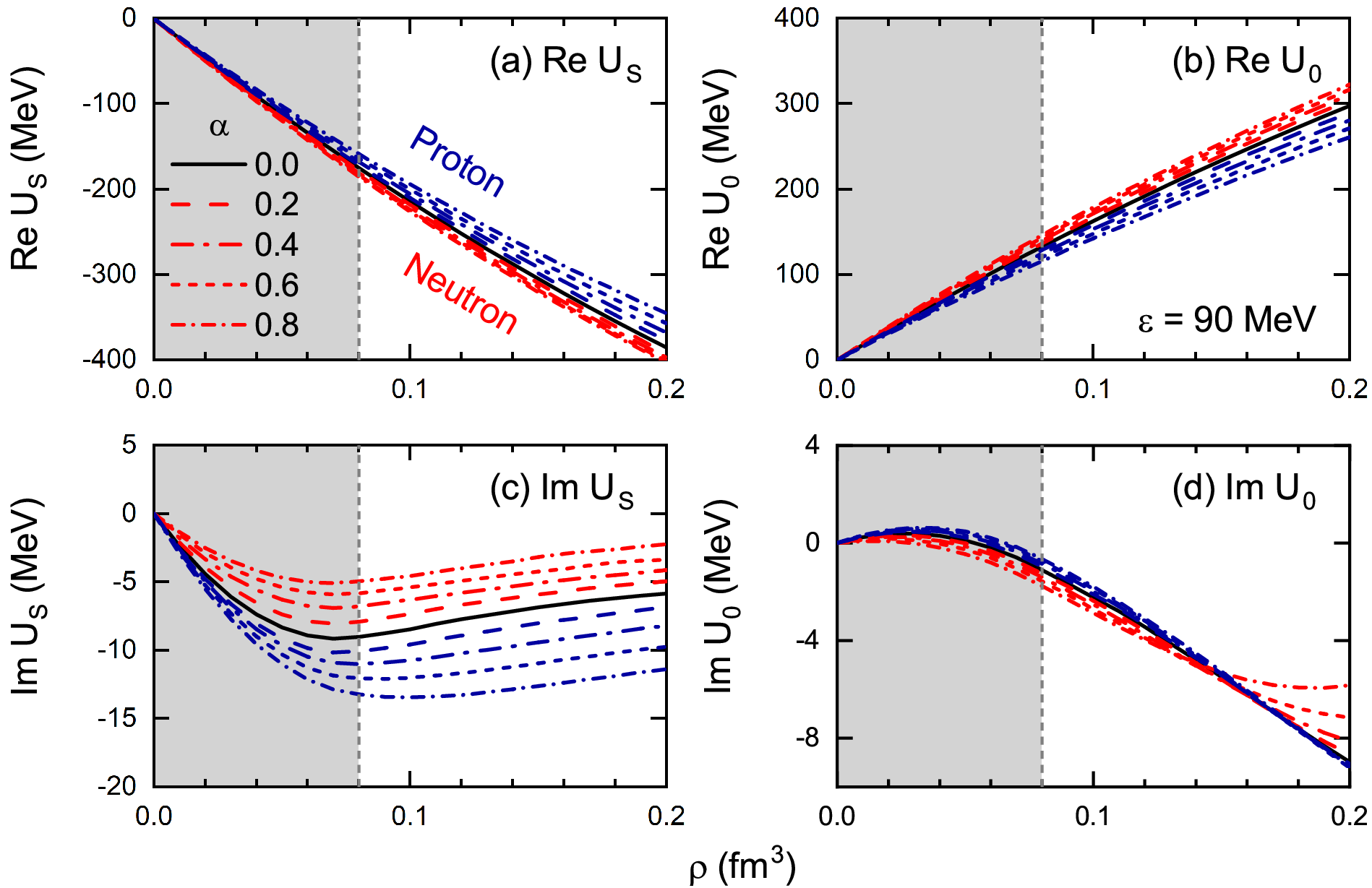}
  \caption{(Color online) The real and imaginary parts of the scalar $U_{S,\tau}$ and vector potential $U_{0,\tau}$ as functions of density with the asymmetry parameter $\alpha$ ranging from 0 to 0.8. The theoretical results in shaded areas on the left of $\rho=0.08\ \text{fm}^{-3}$ are obtained by quadratic extrapolation. The single-particle energy $\varepsilon$ is chosen at 90 MeV. }
  \label{label-Fig3}
\end{figure}

To construct the microscopic optical potential for nucleon-nucleus scattering with the ILDA, the scalar and vector potentials in symmetric and asymmetric nuclear matter at various densities covering the realistic cases of finite nuclei ($0\leq\rho\leq0.16\ \text{fm}^{-3}$) are needed.
However, it is well known that the RBHF calculations are not reliable for density $\rho<0.08\ \text{fm}^{-3}$, due to the cluster effects in low-density nuclear matter.
Therefore, it is necessary to extrapolate the results of single-particle potentials to these low densities.
For CTOM~\cite{XuRR-2016-PRC-94-034606}, the real parts of the single-particle potentials at $\rho=0.04\ \text{fm}^{-3}$ and the imaginary parts of that at $\rho=0.04$ and $0.06\ \text{fm}^{-3}$ are adjusted to achieve the best description of the selected experimental data.
Based on the optimized values at the auxiliary density points, the natural constraints that single-particle potentials vanish at $\rho=0$, and the theoretical calculations at $\rho>0.08\ \text{fm}^{-3}$, the polynomial fittings are employed to derive the single-particle potentials in the full density space.
In this work, to avoid introducing free parameters, a quadratic extrapolation $U_i(\rho) = a_i\rho^2 + b_i\rho + c_i\ (i=S,0)$ is adopted for the low-density region $\rho\leq0.08\ \text{fm}^{-3}$.
For each isospin asymmetry $\alpha$, three corresponding coefficients $a_i, b_i$, and $c_i$ are determined uniquely by the natural constraint that $U_i(\rho=0)=0$ together with the continuities of the single-particle potentials themselves and their first derivatives at the $0.08\ \text{fm}^{-3}$.
In practice, the quadratic extrapolation is used only to get $U_S$ and $U_0$ in the low-density region with step $\Delta \rho=0.01\ \mathrm{fm}^{-3}$. Afterwards, the single-particle potentials for nuclear matter at a given density $\rho$ in the full density region $0\leq\rho\leq0.20\ \text{fm}^{-3}$ are obtained with the quadratic Lagrange interpolation.

In Fig.~\ref{label-Fig3}, we show the real and imaginary parts of the scalar potential $U_{S,\tau}$ and vector potential $U_{0,\tau}$ as functions of density with $\alpha$ ranging from 0 to 0.8. 
The single-particle energy $\varepsilon$ is chosen at 90 MeV, as in Ref.~\cite{XuRR-2016-PRC-94-034606}.
The theoretical results in shaded regions on the left of $\rho=0.08\ \text{fm}^{-3}$ are obtained with the abovementioned scheme.
It can be seen that the quadratic extrapolations for $\text{Re}U_S$ and $\text{Re}U_0$ are very close to the linear extrapolation.
For $\text{Im}U_S$ and $\text{Im}U_0$, the quadratic extrapolations are also plausible.


\begin{figure}[htbp]
  \centering
  \includegraphics[width=12.0cm]{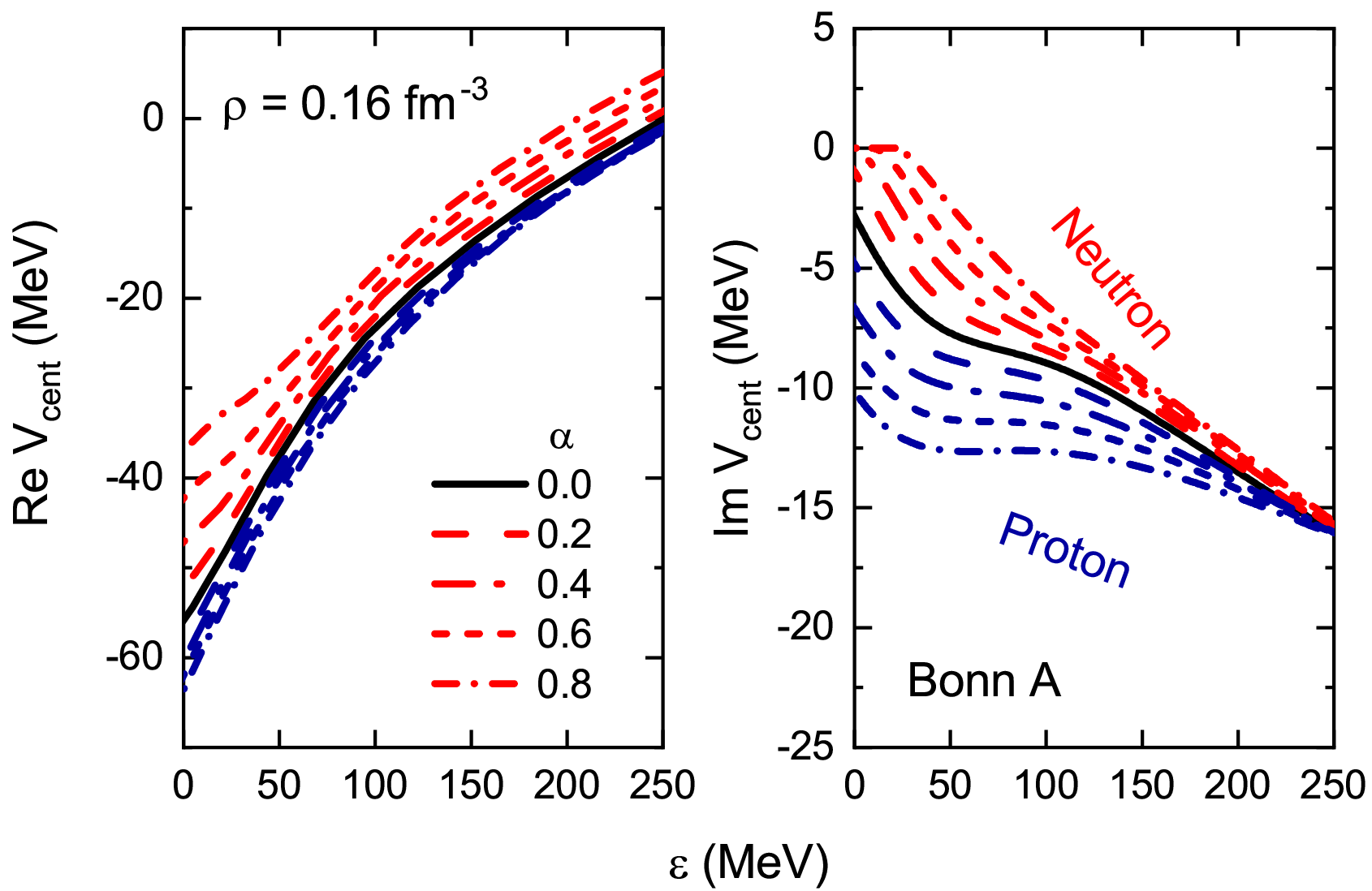}
  \caption{(Color online) The real and imaginary parts of the central potential $V_{\text{cent}}$ as functions of the single-particle energy $\varepsilon$ with the asymmetry parameter $\alpha$ ranging from 0 to 0.8. The density is chosen at $\rho=0.16\ \text{fm}^{-3}$.}
  \label{label-Fig4}
\end{figure}

Starting from the scalar and vector potentials, the real and imaginary parts of the central potential $V_{\text{cent}}$ in Eq.~\eqref{eq:cent} are calculated. They are shown in Fig.~\ref{label-Fig4} as functions of the single-particle energy $\varepsilon$ with the isospin asymmetry parameter $\alpha$ ranging from 0 to 0.8. The density is chosen at $\rho=0.16\ \text{fm}^{-3}$.
It is found that, with the increase of asymmetry parameter, the depths of $\text{Re}V_{\text{cent}}$ and $\text{Im}V_{\text{cent}}$ for the proton increase, while the results for the neutron decrease.
With the increase of incident energy, the depths of $\text{Re}V_{\text{cent}}$ for both proton and neutron decrease, while the results for $\text{Im}V_{\text{cent}}$ are opposite.


\begin{figure}[htbp]
  \centering
  \includegraphics[width=12.0cm]{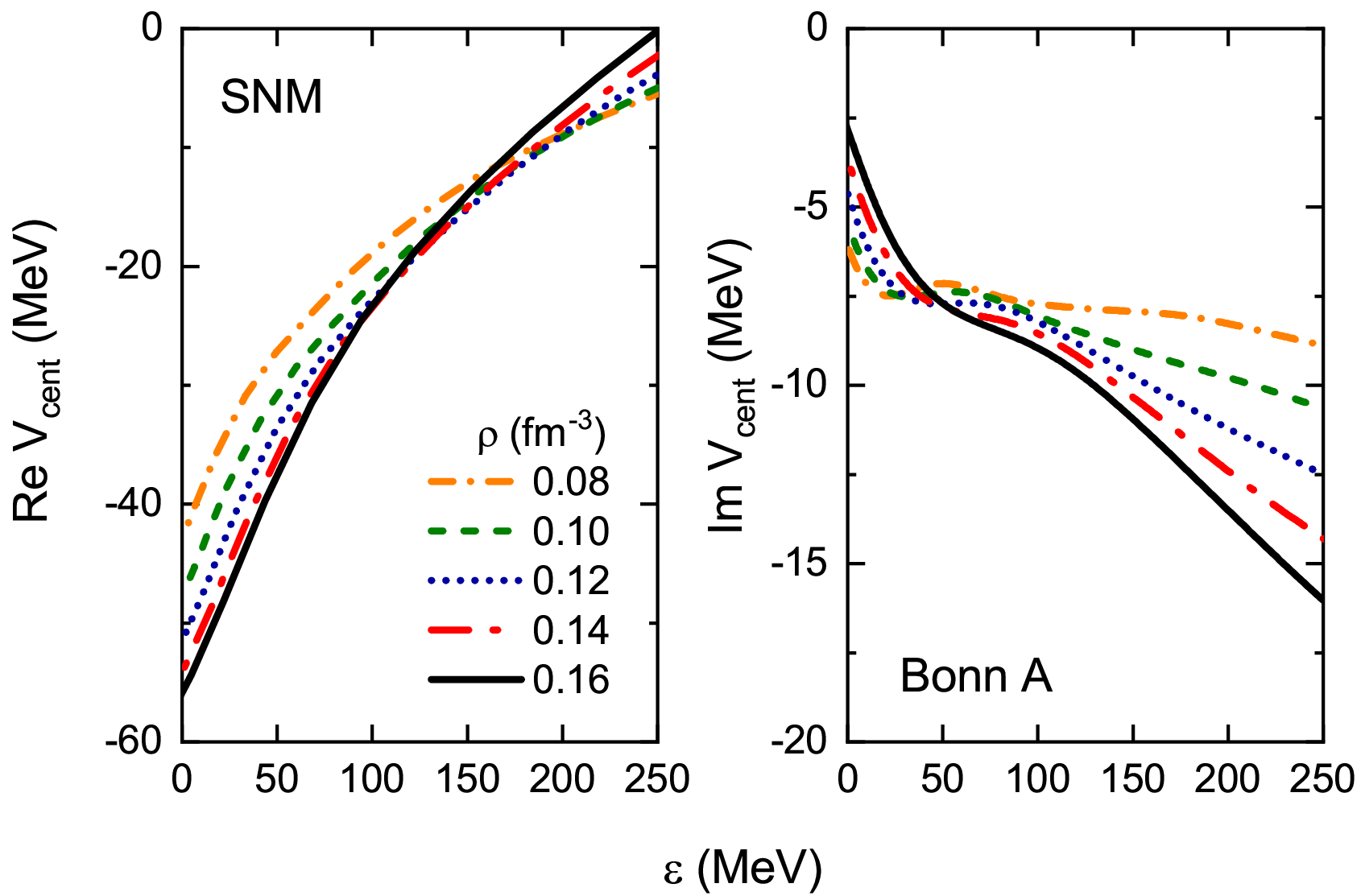}
  \caption{(Color online) The real and imaginary parts of the central potential $V_{\text{cent}}$ as functions of single-particle energy $\varepsilon$ with the density $\rho$ ranging from 0.08 to 0.16 $\text{fm}^{-3}$. The asymmetry parameter is chosen at $\alpha=0$.}
  \label{label-Fig5}
\end{figure}

In Fig.~\ref{label-Fig5}, the real and imaginary parts of the central potential $V_{\text{cent}}$ are shown as functions of single-particle energy $\varepsilon$ with the density ranging from 0.08 to 0.16 $\text{fm}^{-3}$. The asymmetry parameter is chosen at $\alpha=0$.
At energy $\varepsilon \approx 150$\ MeV, a crossing of curves for different densities for $\text{Re}V_{\text{cent}}$ is found, which implies that the so-called wine-bottle-bottom shape~\cite{1981-Arnold-PhysRevC.23.1949} appears at the surface of the target nuclei.
For energy below 50 MeV, the absolute value of $\text{Im}V_{\text{cent}}$ reaches its maximum at lower densities, which indicates a surface absorption.
This is in contrast to higher energies, where the absolute value of $\text{Im}V_{\text{cent}}$ reaches its maximum at higher densities, indicating a volume absorption.

\subsection{Optical potential for proton-nucleus scattering}


\begin{figure}[htbp]
  \centering
  \includegraphics[width=16.0cm]{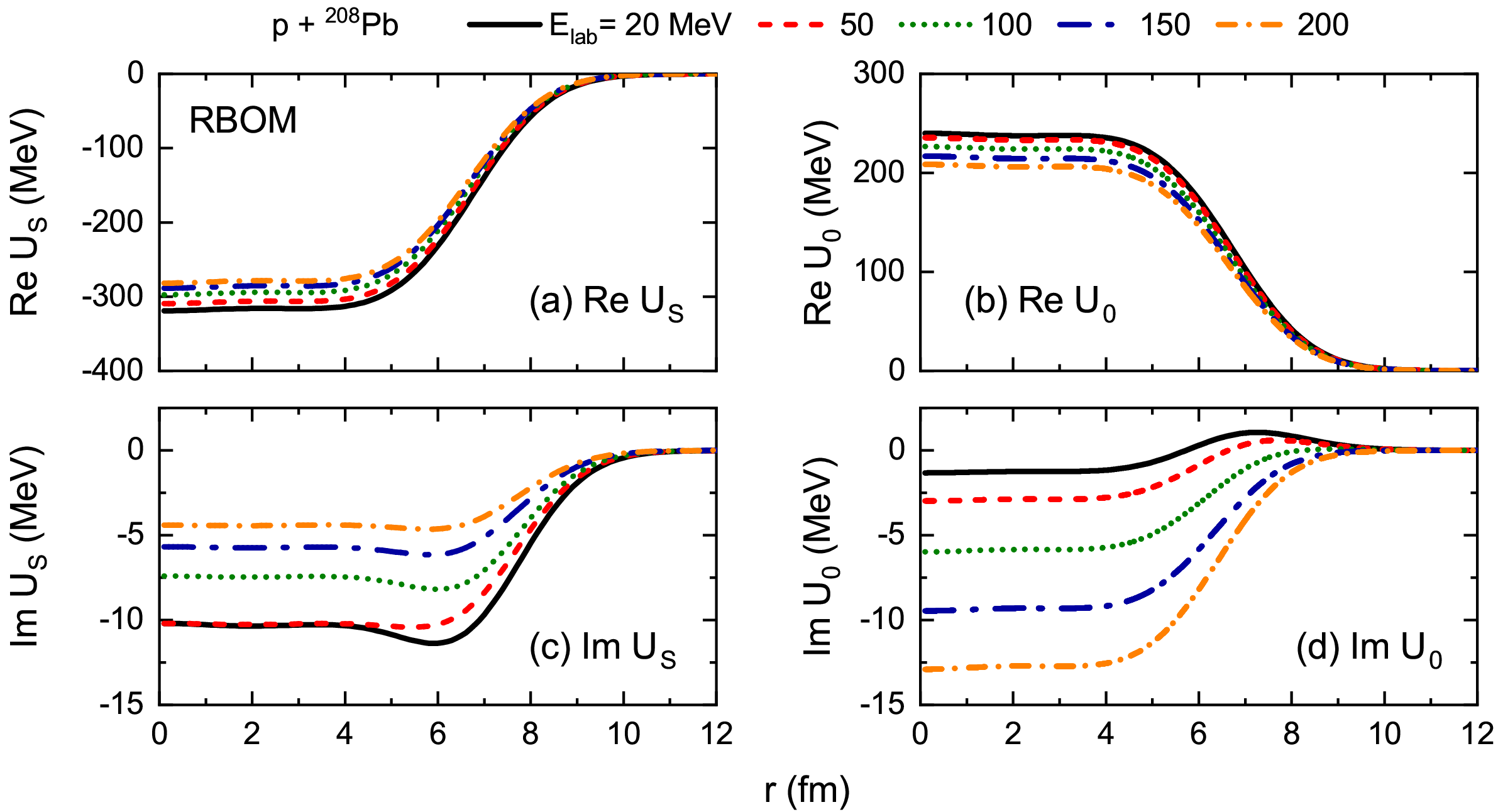}
  \caption{(Color online) The real and imaginary parts of the scalar $U_{S}$ and vector $U_0$ potential as functions of the radial coordinate for $p+\prescript{208}{}{\text{Pb}}$ with $E_{\text{lab}}$ ranging from 20 to 200 MeV.}
  \label{label-Fig6}
\end{figure}

The real and imaginary parts of the scalar and vector components of the single-particle potentials are then used to evaluate the corresponding microscopic optical potential using the ILDA.
The recoil effects are presumably small for heavy targets, but this may be different for lighter targets.
For consistency, the recoil corrections for all targets are considered as in Ref.~\cite{Cooper-1993-PhysRevC.47.297}, where the Cooper-Jennings recoil factors~\cite{1988-Cooper-Nucl.Phys.A} in both scalar and vector potentials are introduced
\begin{equation}
	R_S = M_t/\sqrt{s}, \qquad R_V = E_t/\sqrt{s}.
\end{equation}
Here $\sqrt{s}$ is the c.m. energy of $p+A$ system, and $E_t$ is the total c.m. energy of the target.
For $\prescript{208}{}{\text{Pb}}$, these factors are close to unity, while, for $p+\prescript{16}{}{\text{O}}$ scattering at 400 MeV, $R_S$ is 0.92.

In Fig.~\ref{label-Fig6}, we show the real and imaginary parts of the scalar $U_{S}$ and vector $U_0$ potentials as functions of the radial coordinate for $p+\prescript{208}{}{\text{Pb}}$ with $E_{\text{lab}}$ ranging from 20 to 200 MeV.
Their radial profiles resemble the nuclear density, with minor exceptions found for the imaginary parts of $U_S$ and $U_0$ at $E_{\text{lab}}=20$\ MeV.
With the increase of the incident energy, the depths of Re$U_{S}$, Re$U_0$, and Im$U_{S}$ decrease, while the depth of Im$U_0$ increases. 
We find that, the real and imaginary parts have different geometries, especially at 20 MeV for either $U_S$ or $U_0$.


\begin{figure}[htbp]
  \centering
  \includegraphics[width=16.0cm]{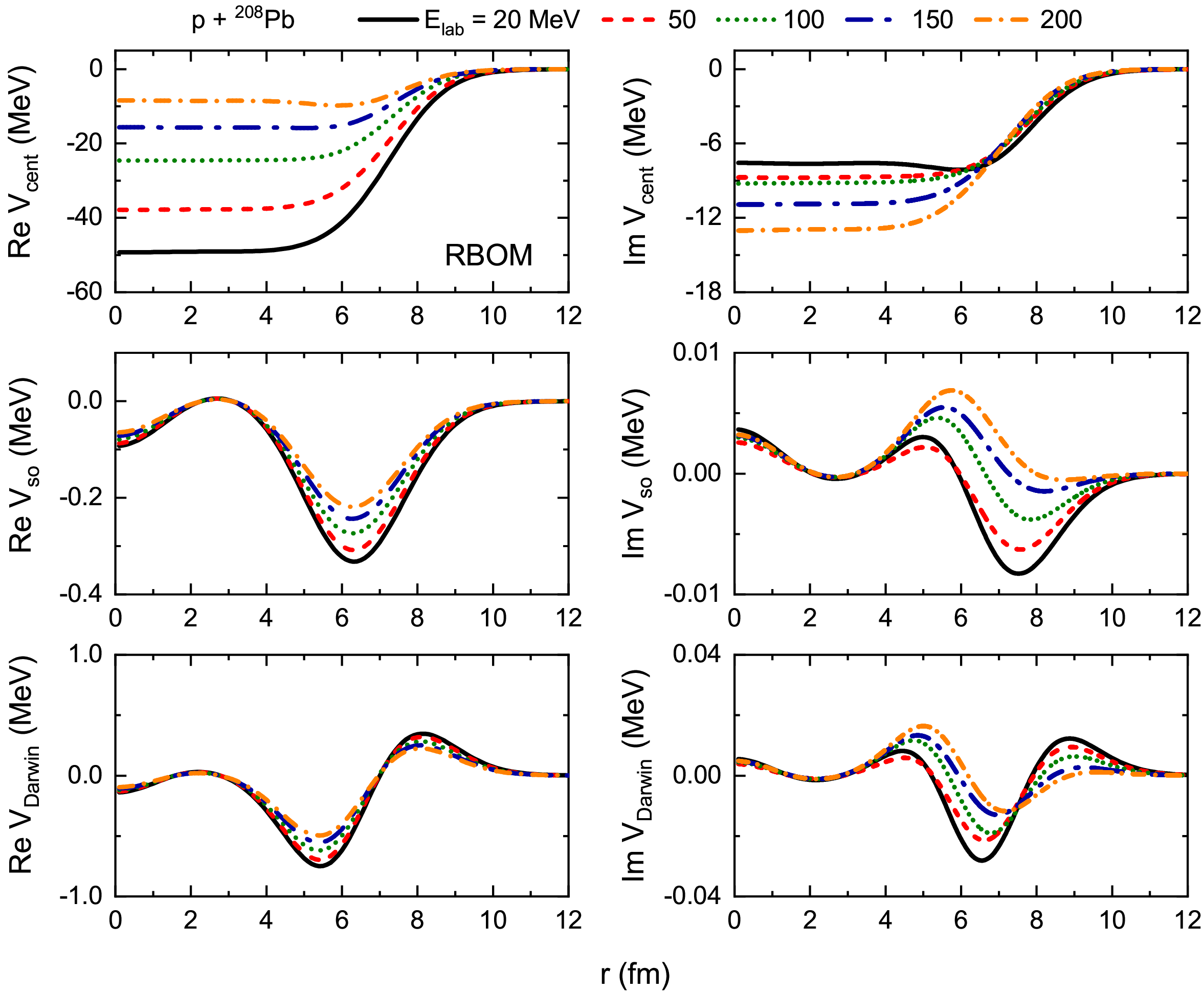}
  \caption{(Color online) The real and imaginary parts of the central term $V_{\text{cent}}$, spin-orbit term $V_{\text{so}}$, and Darwin term $V_{\text{Darwin}}$ of the optical potential as functions of the radial coordinate for $p+\prescript{208}{}{\text{Pb}}$ with $E_{\text{lab}}$ ranging from 20 to 200 MeV.}
  \label{label-Fig7}
\end{figure}

Different components of the optical potential for proton-nucleus scattering can be obtained with the ILDA and the scalar $U_{S}$ and vector $U_0$ potentials.
The real and imaginary parts of the central term $V_{\text{cent}}$, spin-orbit term $V_{\text{so}}$, and Darwin term $V_{\text{Darwin}}$ of the optical potential are shown in Fig.~\ref{label-Fig7}, as functions of the radial coordinate for $p+\prescript{208}{}{\text{Pb}}$ with the $E_{\text{lab}}$ ranging from 20 to 200 MeV.
For the central term, as the energy increases, the depths of the real potentials decrease while those of the imaginary parts increase. 
It is noticed that the location where Im$V_{\text{cent}}$ reaches its maximum magnitude changes from $r\simeq 7$ fm at $E_{\text{lab}}=20$\ MeV to $r\simeq 0$ fm at $E_{\text{lab}}=200$\ MeV, revealing a transition from surface absorption to volume absorption.
It can also be observed that the spin-orbit and Darwin terms are marginal compared to the central terms.
These behaviors are consistent with the phenomenological optical potentials~\cite{1987-Cooper-PhysRevC.36.2170, Hama-1990-PhysRevC.41.2737, Cooper-1993-PhysRevC.47.297}.


\begin{figure}[htbp]
  \centering
  \includegraphics[width=16.0cm]{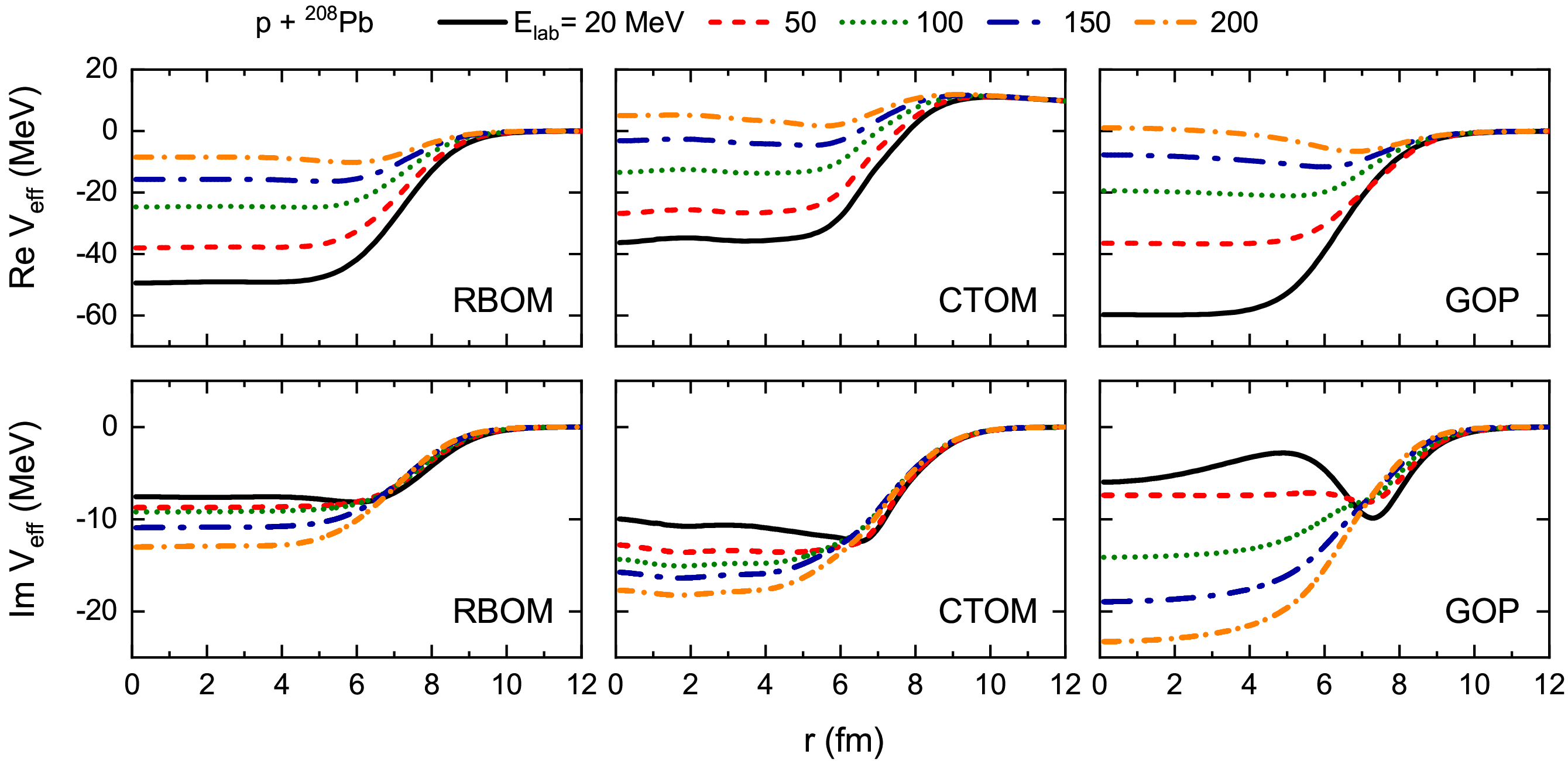}
  \caption{(Color online) The real and imaginary parts of the central potentials $V_{\text{eff}}=V_{\text{cent}}+V_{\text{Darwin}}$ as functions of the radial coordinate for $p+\prescript{208}{}{\text{Pb}}$ with $E_{\text{lab}}$ ranging from 20 to 200 MeV, in comparison with the results from CTOM and GOP.}
  \label{label-Fig8}
\end{figure}

In Fig.~\ref{label-Fig8}, we compare our central potentials $V_{\text{eff}}=V_{\text{cent}}+V_{\text{Darwin}}$ to the ones obtained with the microscopic CTOM~\cite{XuRR-2016-PRC-94-034606} and the phenomenological GOP~\cite{2009-Cooper-PhysRevC.80.034605}.
Five incident energies from 20 to 200 MeV are considered. 
The RBOM potential and the CTOM potential are both derived from the RBHF calculations for infinite nuclear matter with the ILDA. 
Qualitatively, despite of the differences, including the treatments of NESs, the low-density extrapolations, and the target densities, the real and imaginary parts of the central potentials are in good agreement between the two relativistic microscopic optical potentials. 
Quantitatively, the depths of the real central potentials obtained with the CTOM potential are slightly lower than in the case of RBOM, and the response of the depths of the imaginary parts with respect to the incident energy is more evident than with RBOM.

Compared to the cases of microscopic RBOM and CTOM, the depths of the real and imaginary potentials obtained with the phenomenological GOP show a much stronger energy dependence.
Besides, the oscillating behavior in the profile of $\text{Im}V_{\text{cent}}$ from $r=4$ to 8 fm for GOP at incident energy $E_{\text{lab}}=20$ MeV is not observed, neither in RBOM nor in CTOM.
Since the parameters of GOP are determined by fitting to proton-nucleus scattering data with incident energy no smaller than 20 MeV, the unusual oscillating behavior found by GOP might indicate its poor applicability for smaller incident energies.


\begin{figure}[htbp]
  \centering
  \includegraphics[width=16.0cm]{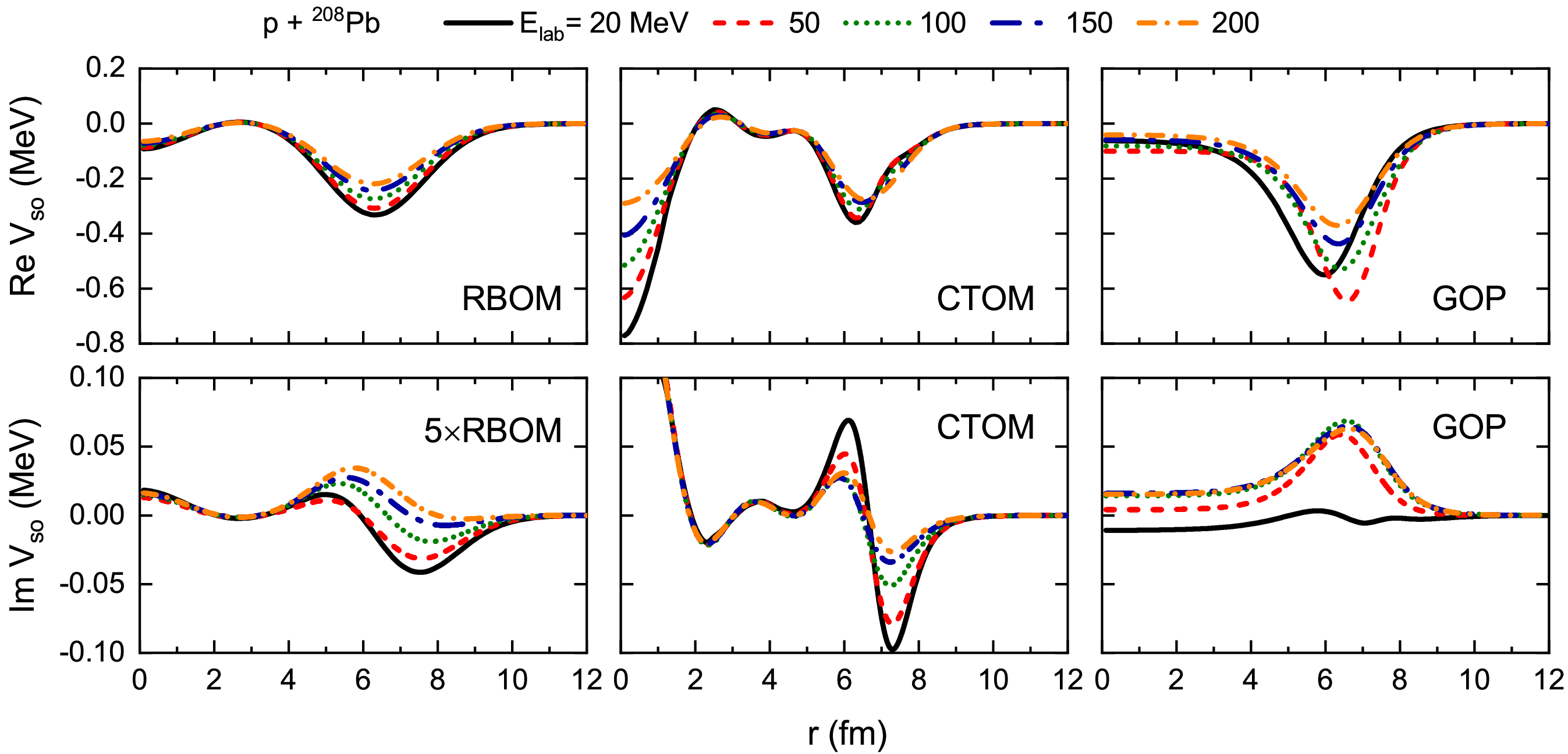}
  \caption{(Color online) The real and imaginary parts of the spin-orbit potentials as functions of the radial coordinate for $p+\prescript{208}{}{\text{Pb}}$ with the $E_{\text{lab}}$ ranging from 20 to 200 MeV, in comparison with the results from CTOM and GOP. 
  The results for $\text{Im}V_{\text{so}}$ from RBOM are amplified by a factor of 5.}
  \label{label-Fig9}
\end{figure}

Figure \ref{label-Fig9} compares the spin-orbit potentials among the RBOM, the CTOM, and the phenomenological GOP.
Overall, the discrepancy is much more pronounced than the central potentials as shown in Fig.~\ref{label-Fig8}.
Notice that the results for $\text{Im}V_{\text{so}}$ from RBOM are amplified by a factor of 5.
In addition to the magnitudes, apparent discrepancies in the spin-orbit potentials between the RBOM and the CTOM are found.
By replacing the density distribution of $\prescript{208}{}{\text{Pb}}$ in this work to that used in CTOM, which is obtained by the Hartree-Fock Bogoliubov approach with Gogny D1S force~\cite{2007-Hilaire-EPJA}, the discrepancies in the spin-orbit potentials are barely reduced.
This implies that the single-particle potentials $U_S, U_0$ in RBOM are different from those in CTOM, indicating the importance of the NESs.

Compared to our RBOM, both the magnitudes and the energy dependence of the spin-orbit potentials from the GOP are stronger, especially for the imaginary part. 
For the profiles of spin-orbit potentials from the GOP, the results at $E_{\text{lab}}=20$\ MeV are not consistent with the other cases with larger incident energies. 
This is quite different from the case found with the RBOM potential, where continuous changes of the profiles with respect to the incident energies are obtained.
Again, the special behaviors for GOP at small incident energy might be related to the fact the 20 MeV is on the lower edge of the incident energies in the fitting procedure.

\subsection{Scattering quantities from the RBOM potential}


\begin{figure}[htbp]
  \centering
  \includegraphics[width=16.0cm]{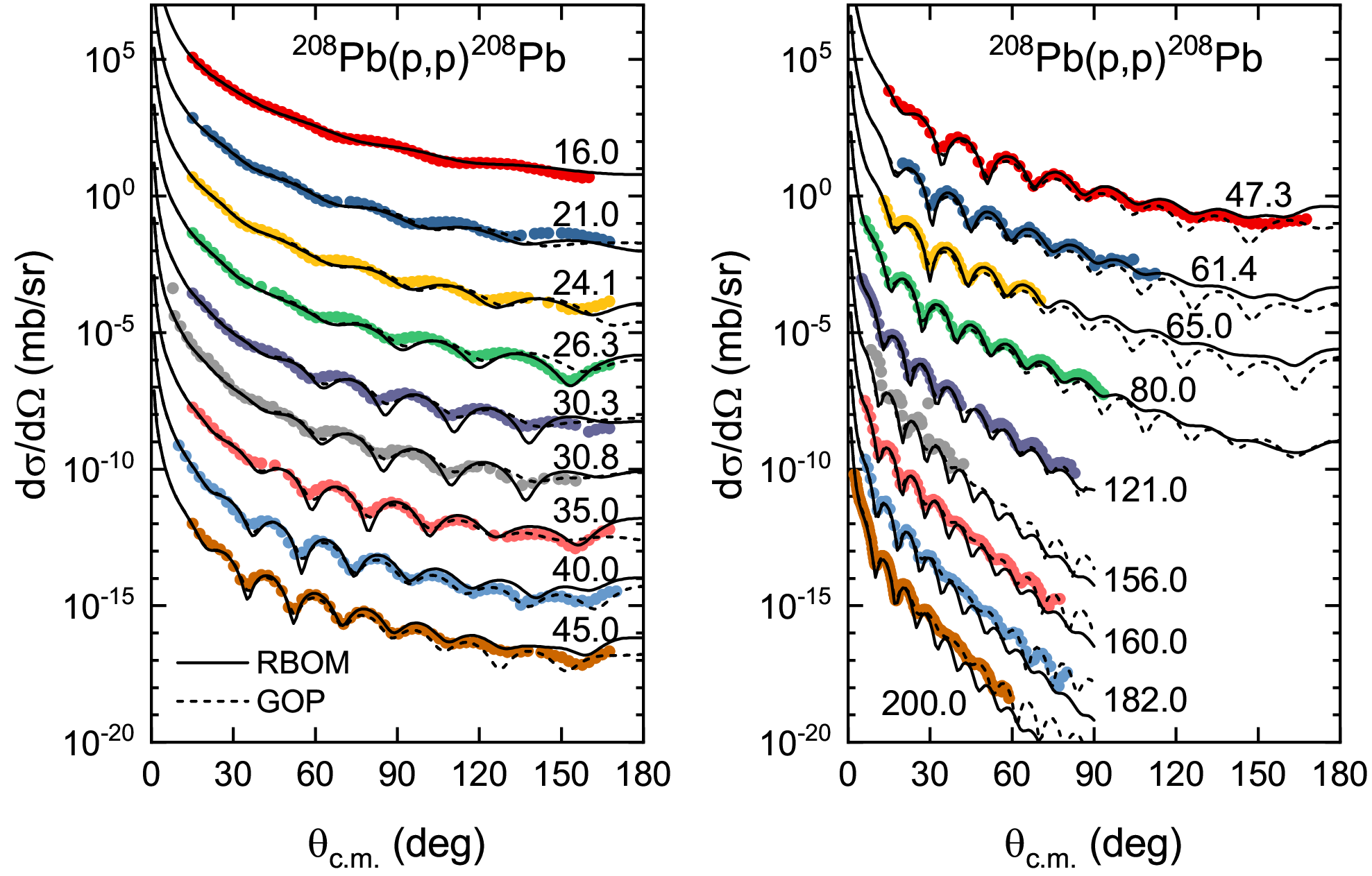}
  \caption{(Color online) The elastic scattering differential cross section as a function of scattering angle in the c.m. frame, calculated by the RBHF theory in combination with the ILDA, for the scattering of protons from $\prescript{208}{}{\text{Pb}}$ with energy ranging from 16.0 to 182.0 MeV. The corresponding experimental data available are shown. For comparison, the results from GOP are also shown.}
  \label{label-Fig10}
\end{figure}

In this subsection, the present microscopic optical potential RBOM is assessed through the predictions of the experimental observables of proton-nucleus scattering for five different targets $\prescript{208}{}{\text{Pb}}$, $\prescript{120}{}{\text{Sn}}$, $\prescript{90}{}{\text{Zr}}$, $\prescript{48}{}{\text{Ca}}$, and $\prescript{40}{}{\text{Ca}}$ with incident energies below 200 MeV. 
The elastic scattering differential cross sections calculated with RBOM for proton scattering off $\prescript{208}{}{\text{Pb}}$ are given in Fig.~\ref{label-Fig10}. 
The corresponding experimental data and the results calculated with GOP~\cite{2009-Cooper-PhysRevC.80.034605} are also plotted in the same figure.
The experimental data adopted in our analysis are all from the EXFOR library~\cite{[https://www-nds.iaea.org/exfor/]url-EXFOR}, which is a comprehensive database that gathers the world's nuclear reaction measurements. 
Since GOP is restricted down to 20 MeV, we only show the results calculated with GOP for larger incident energies.
Notice that the curves and data points at the top are true values, while the others are offset by factors of $10^{-2}$, $10^{-4}$, etc.

Overall, our results are in good agreement with the experimental data, especially for the scattering of protons with incident energy around 80 MeV.
Slight overestimates of the differential cross sections with smaller incident energies and underestimates of those with larger incident energies are found.
For large angle scattering with incident energies in the interval of 60-80 MeV, our angular distributions are flatter than that of GOP and are more likely to be favored by experimental data, as inferred from existing data for smaller angles.

\begin{figure}[htbp]
  \centering
  \includegraphics[width=16.0cm]{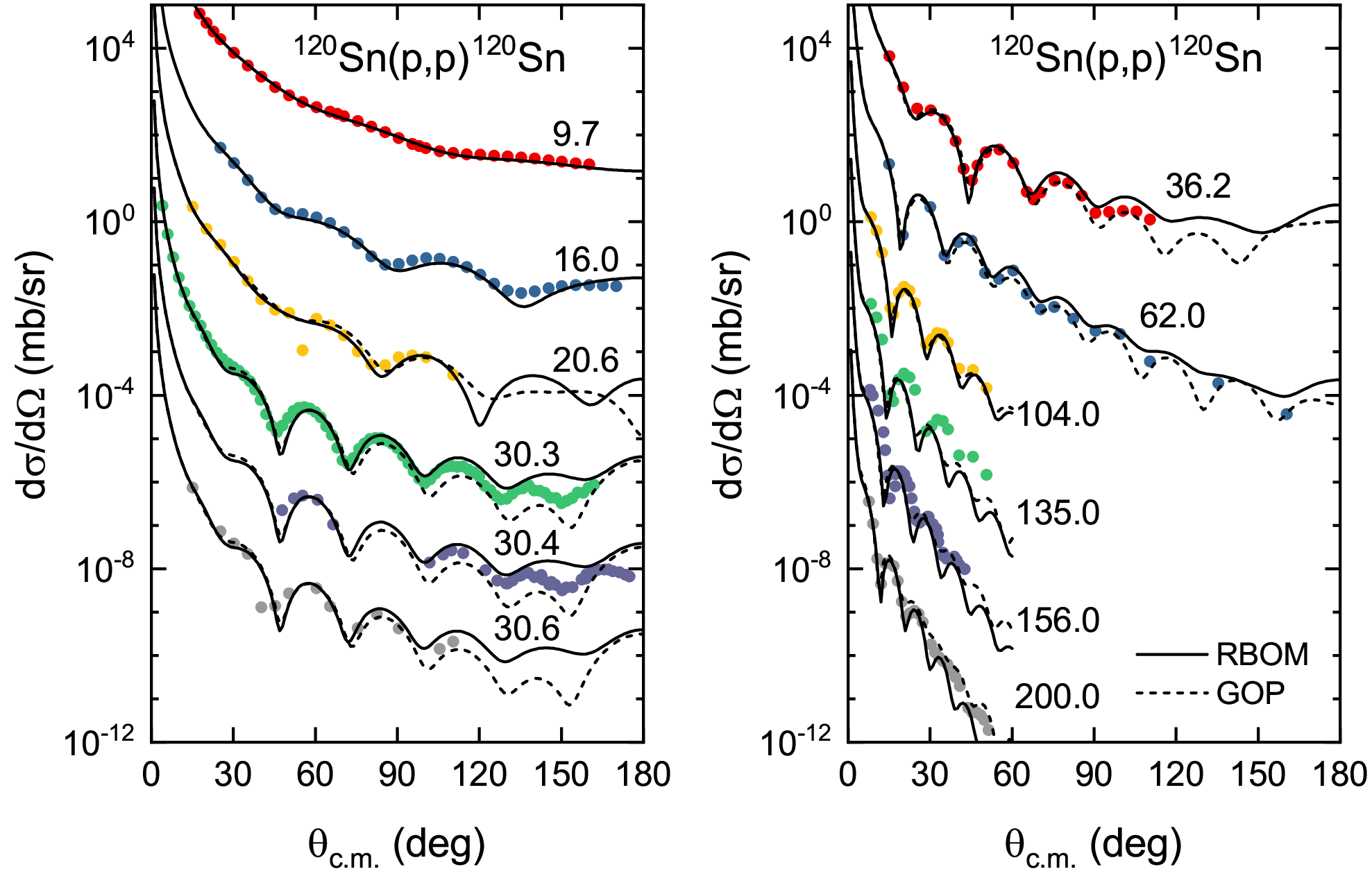}
  \caption{(Color online) Similar to Fig.~\ref{label-Fig10}, but for the scattering of protons from $\prescript{120}{}{\text{Sn}}$ with energy ranging from 9.7 to 200.0 MeV. }
  \label{label-Fig11}
\end{figure}

\begin{figure}[htbp]
  \centering
  \includegraphics[width=16.0cm]{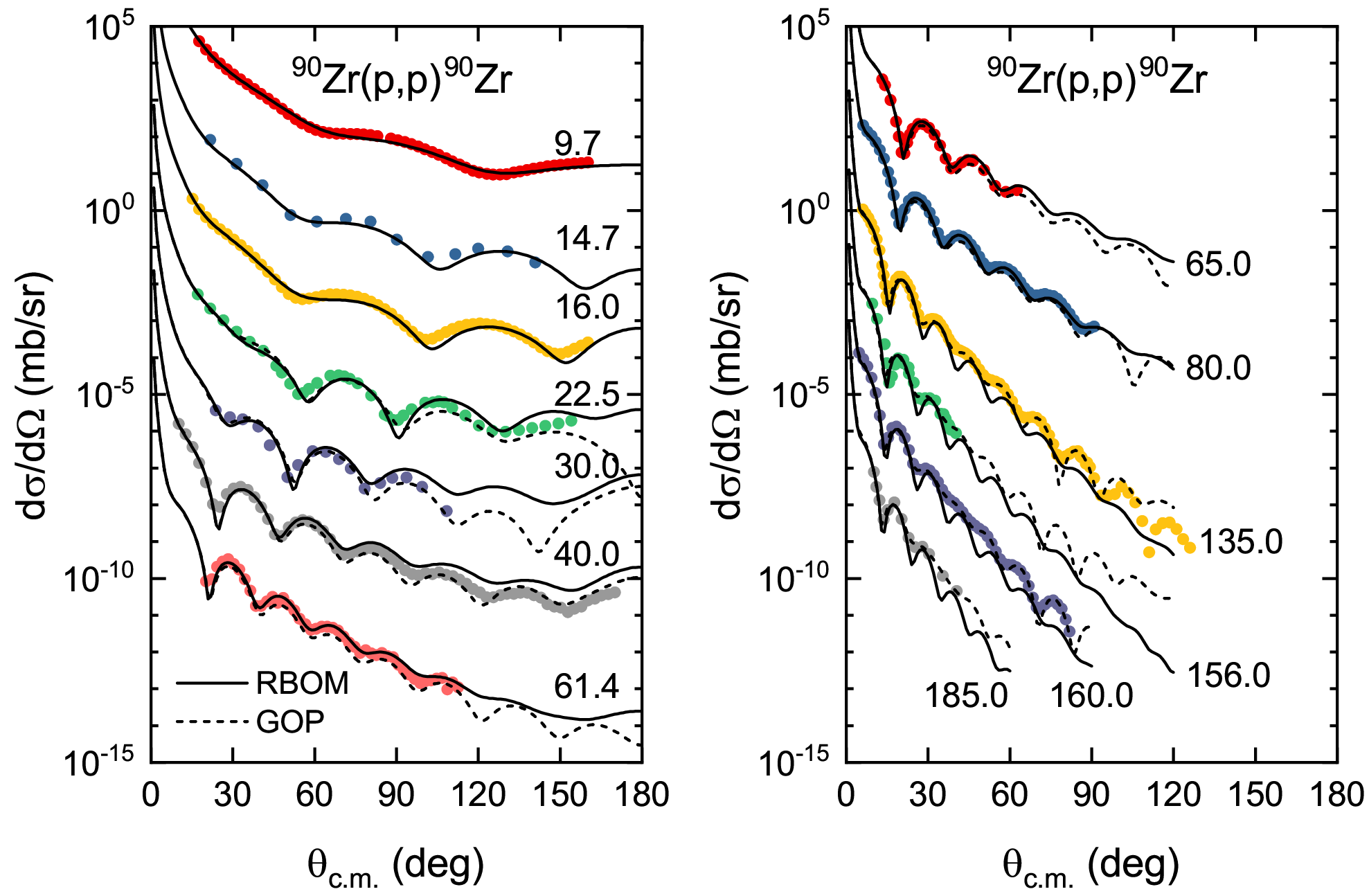}
  \caption{(Color online) Similar to Fig.~\ref{label-Fig10}, but for the scattering of protons from $\prescript{90}{}{\text{Zr}}$ with energy ranging from 9.7 to 185.0 MeV. }
  \label{label-Fig12}
\end{figure}

\begin{figure}[htbp]
  \centering
  \includegraphics[width=16.0cm]{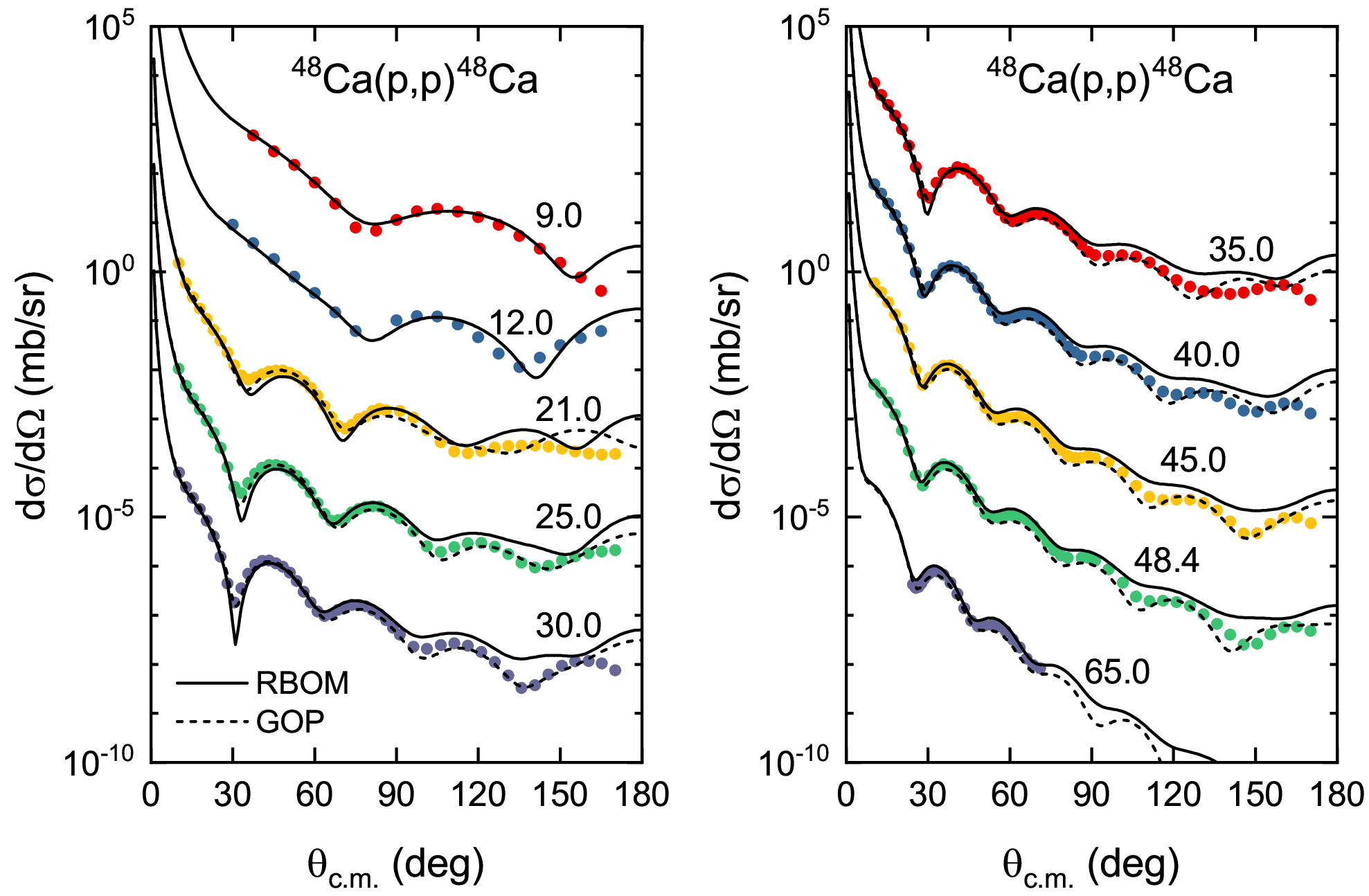}
  \caption{(Color online) Similar to Fig.~\ref{label-Fig10}, but for the scattering of protons from $\prescript{48}{}{\text{Ca}}$ with energy ranging from 9.0 to 65.0 MeV. }
  \label{label-Fig13}
\end{figure}

\begin{figure}[htbp]
  \centering
  \includegraphics[width=16.0cm]{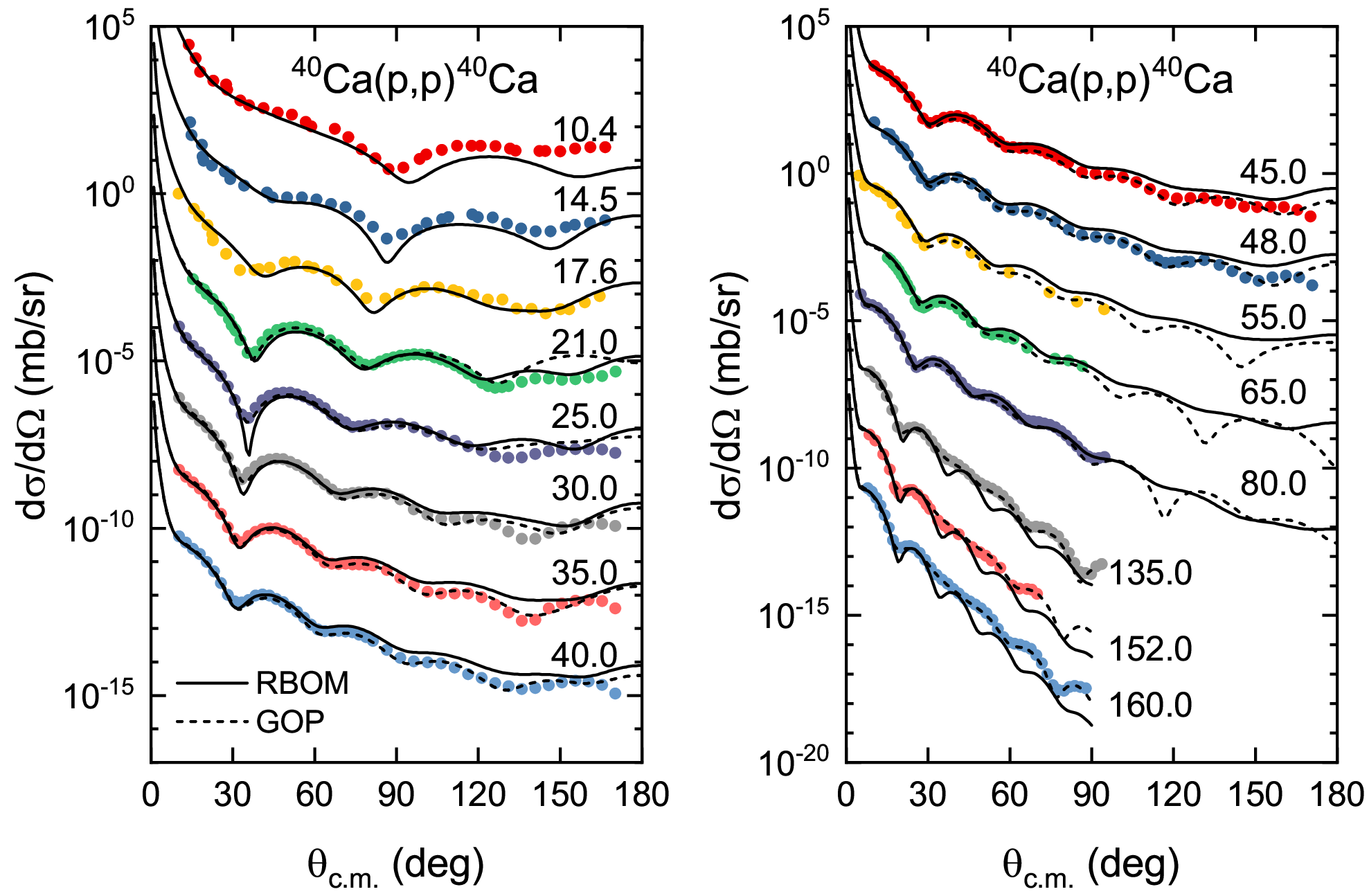}
  \caption{(Color online) Similar to Fig.~\ref{label-Fig10}, but for the scattering of protons from $\prescript{40}{}{\text{Ca}}$ with energy ranging from 10.4 to 160.0 MeV. }
  \label{label-Fig14}
\end{figure}

In addition to $\prescript{208}{}{\text{Pb}}$, we also show the elastic scattering differential cross sections calculated with our optical potential for $\prescript{120}{}{\text{Sn}}$, $\prescript{90}{}{\text{Zr}}$, $\prescript{48}{}{\text{Ca}}$, and $\prescript{40}{}{\text{Ca}}$ in Figs.~\ref{label-Fig11}-\ref{label-Fig14}.
They are compared to the experimental data and phenomenological optical potential GOP results.
Good reproduction of experimental data with incident energy close to 80 MeV is also found.
Considering that there is no free parameter other than the range factor $t$ in the RBOM, this assessment clearly shows the satisfactory performance of the relativistic microscopic optical potential developed in this work.


\begin{figure}[htbp]
  \centering
  \includegraphics[width=16.0cm]{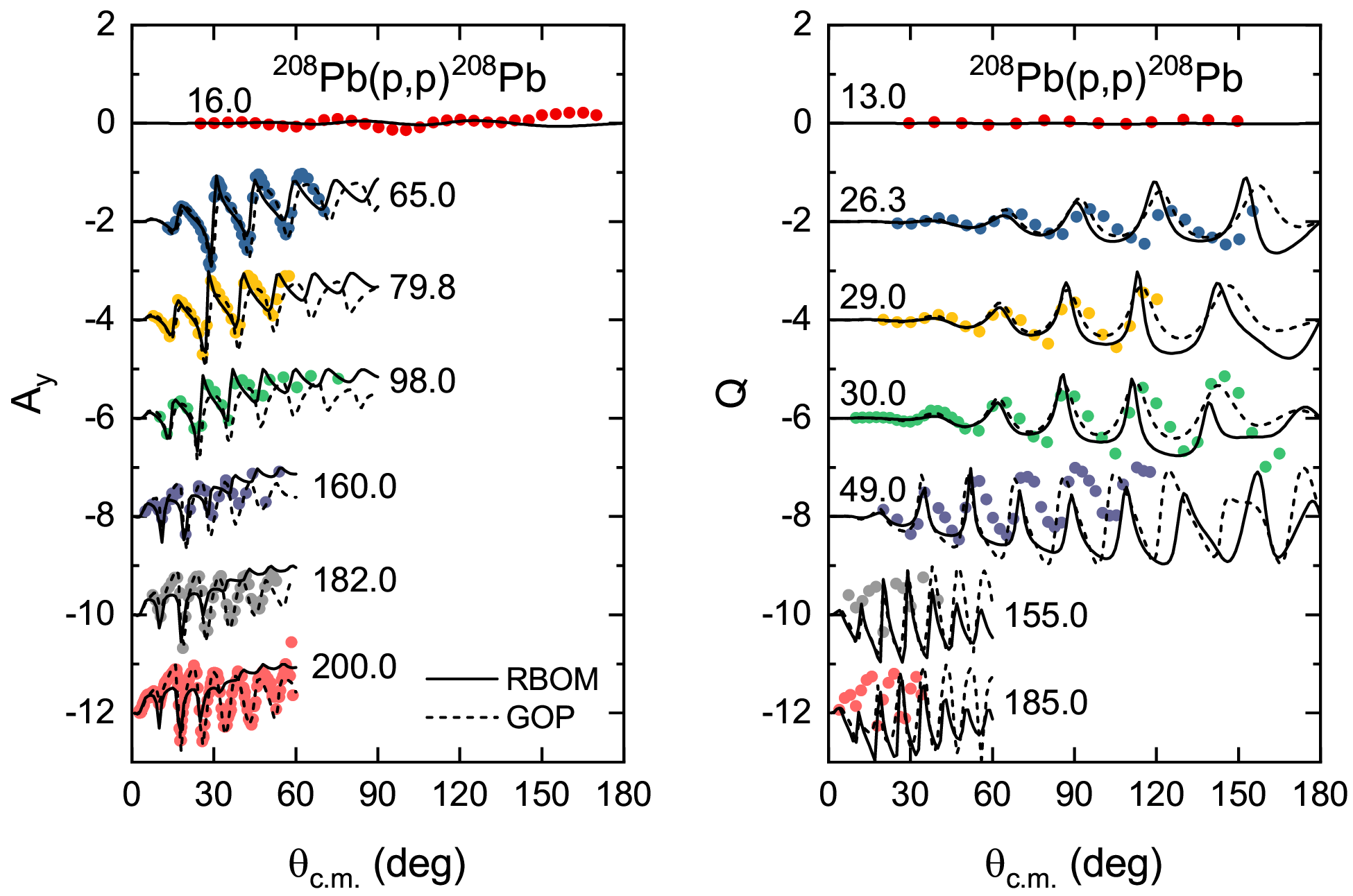}
  \caption{(Color online) The quantities $A_y$ and $Q$ as a function of center of mass angle, calculated by the RBHF theory in combination with the ILDA, for the scattering of protons from $\prescript{208}{}{\text{Pb}}$ with energy ranging from 16.0 to 185.0 MeV. The corresponding experimental data available are shown. For comparison, the results from GOP are also shown.}
  \label{label-Fig15}
\end{figure}

An important feature of the relativistic description of nucleon-nucleus scattering is that the spin-orbit term can be naturally involved without any additional parameter, which is significant for the derivation of the spin observables $A_y(\theta)$ and $Q(\theta)$~\cite{1983-Clark-Phys.Lett.B121.211}.
In Fig.~\ref{label-Fig15}, we show the analyzing power $A_y$ and spin rotation function $Q$ for a proton scattering off $\prescript{208}{}{\text{Pb}}$ calculated with the RBOM potential in this work. 
The curves and data points at the top are true values, while the others are offset by factors of $-2$, $-4$, etc. 
The predicted phases of $A_y$ and $Q$ look well, whereas the amplitudes are not ideal, especially for those with incident energy above 100 MeV.
The results from GOP are also shown for comparison. 
Better agreement is found between our calculation and GOP for smaller energies and angles.


\begin{figure}[htbp]
  \centering
  \includegraphics[width=12.0cm]{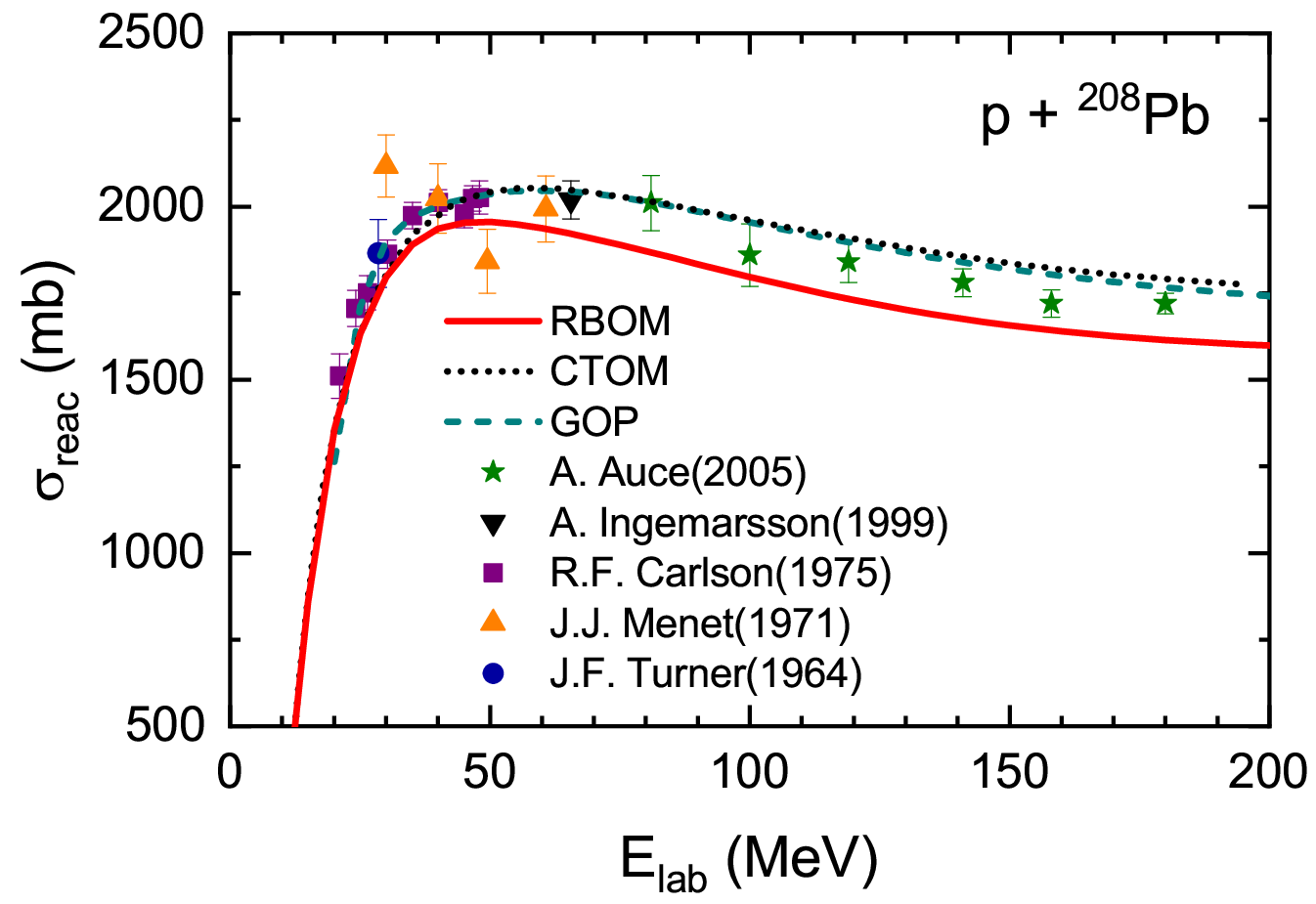}
  \caption{(Color online) The reaction cross section for $p+\prescript{208}{}{\text{Pb}}$ calculated by the RBOM, in comparison with experimental data available and the results obtained by CTOM.}
  \label{label-Fig16}
\end{figure}

In Fig.~\ref{label-Fig16}, we show the reaction cross section for proton scattering off $\prescript{208}{}{\text{Pb}}$ calculated by the RBOM potential.
For comparison, we also show the experimental data available and the results obtained by the CTOM potential and GOP. It is not surprising to find that the results from the CTOM potential and GOP are very close, considering that the experimental data of proton scattering off $\prescript{208}{}{\text{Pb}}$ have been used in the global fitting of GOP as well as in the optimization of CTOM.
For incident energies below 50 MeV, our results are consistent with the experimental data, and are close to the results from CTOM.
For energies above, the reaction cross sections are underestimated by our RBOM potential, while they are overestimated by the CTOM.

\subsection{Performance of the RBOM potential at high incident energies}

As shown in previous subsection, the newly developed RBOM optical potential performs less well with high incident energies around 200 MeV. We stress that no parameter has been adjusted to experimental data. Nevertheless, it is still beneficial to analyze the physical reason for the unsatisfactory performance of the RBOM potential at high incident energies.

\begin{figure}[htbp]
  \centering
  \includegraphics[width=14.5cm]{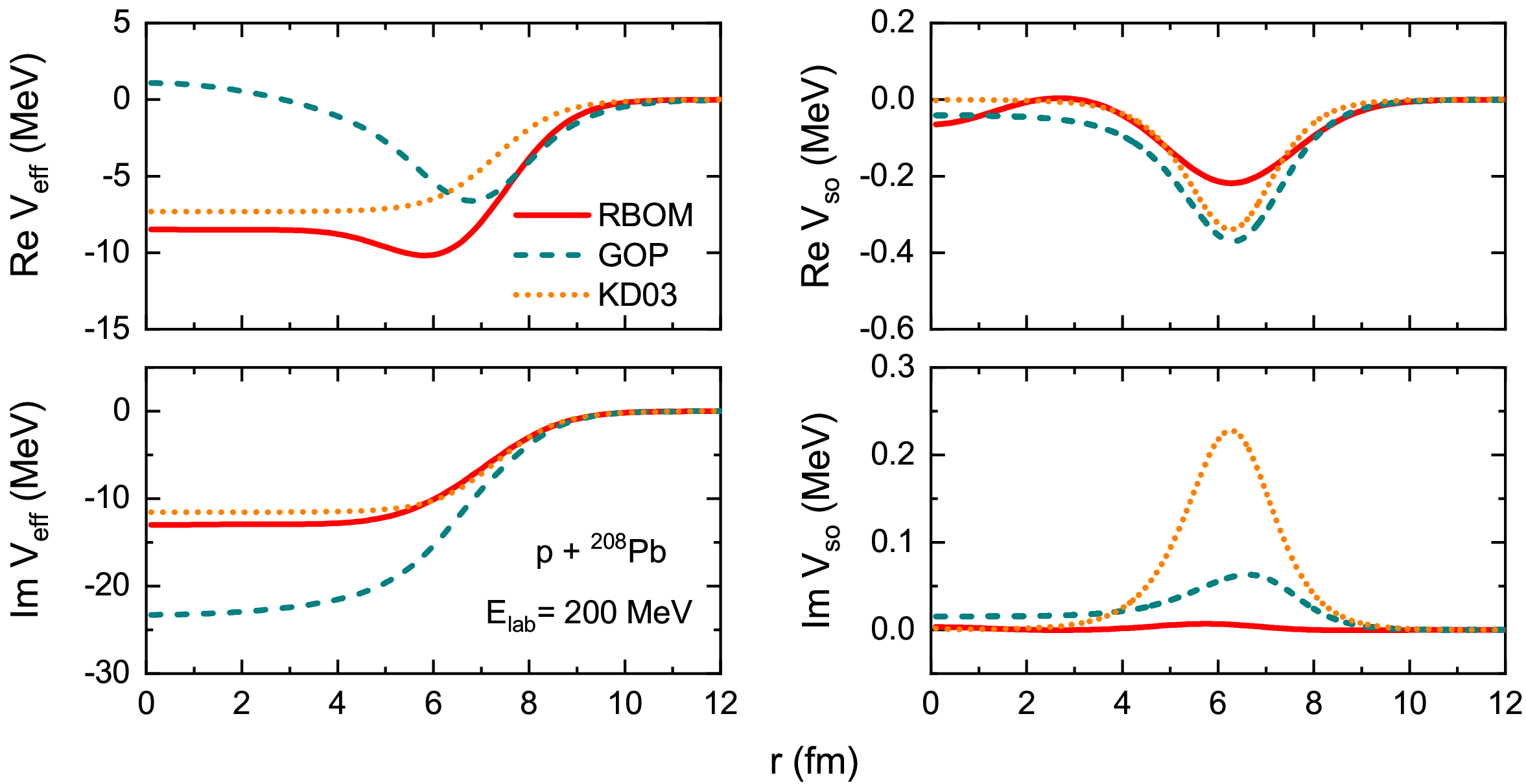}
  \caption{(Color online) The real and imaginary parts of the central potential $V_{\text{eff}}$ as well as the spin-orbit potential $V_{\text{so}}$ as functions of the radial coordinate for $p+\prescript{208}{}{\text{Pb}}$ with $E_{\text{lab}}=200$\ MeV, in comparison with the results from GOP and KD03.}
  \label{label-FigS0}
\end{figure}

Taking proton scattering off $\prescript{208}{}{\text{Pb}}$ with incident energy 200 MeV for an example, in Fig.~\ref{label-FigS0}, we compare the real and imaginary parts of the central potential $V_{\text{eff}}$ as well as the spin-orbit potential $V_{\text{so}}$ of the RBOM potential with that of the phenomenological relativistic optical potential GOP. The results of non-relativistic optical potential KD03 are also included in the comparison. Firstly, although both GOP and KD03 are fitted to experimental data including the cross section angular distributions, their central potentials show evident difference. Secondly, the central potentials from RBOM are very close to those of KD03. The combination of the two points might imply that the underestimation of cross section angular distributions by RBOM is not due to the central potentials. 

For the spin-orbit potentials, as shown in Fig.~\ref{label-FigS0}, the magnitudes of the results from RBOM are found to be smaller than those from both GOP and KD03. Suppressing Im$V_{\text{so}}$ of RBOM gives no apparent changes of cross section angular distributions since they are extremely small, while suppressing Re$V_{\text{so}}$ shows clear difference. In particular, doubling Re$V_{\text{so}}$ leads to results much closer to experimental data at large angles. According to Eq.~\eqref{eq:vso}, the quantity Re$V_{\text{so}}$ is related to energy dependence and the density dependence of single-particle potentials. Considering that the underestimation of angular distributions also holds for other nuclei with significantly different density distributions in comparison to $\prescript{208}{}{\text{Pb}}$, the density dependence of single-particle potentials might not be the main reason behind the underestimation. Therefore, we pay attention to the energy dependence of the single-particle potentials.

\begin{figure}[htbp]
  \centering
  \includegraphics[width=14.5cm]{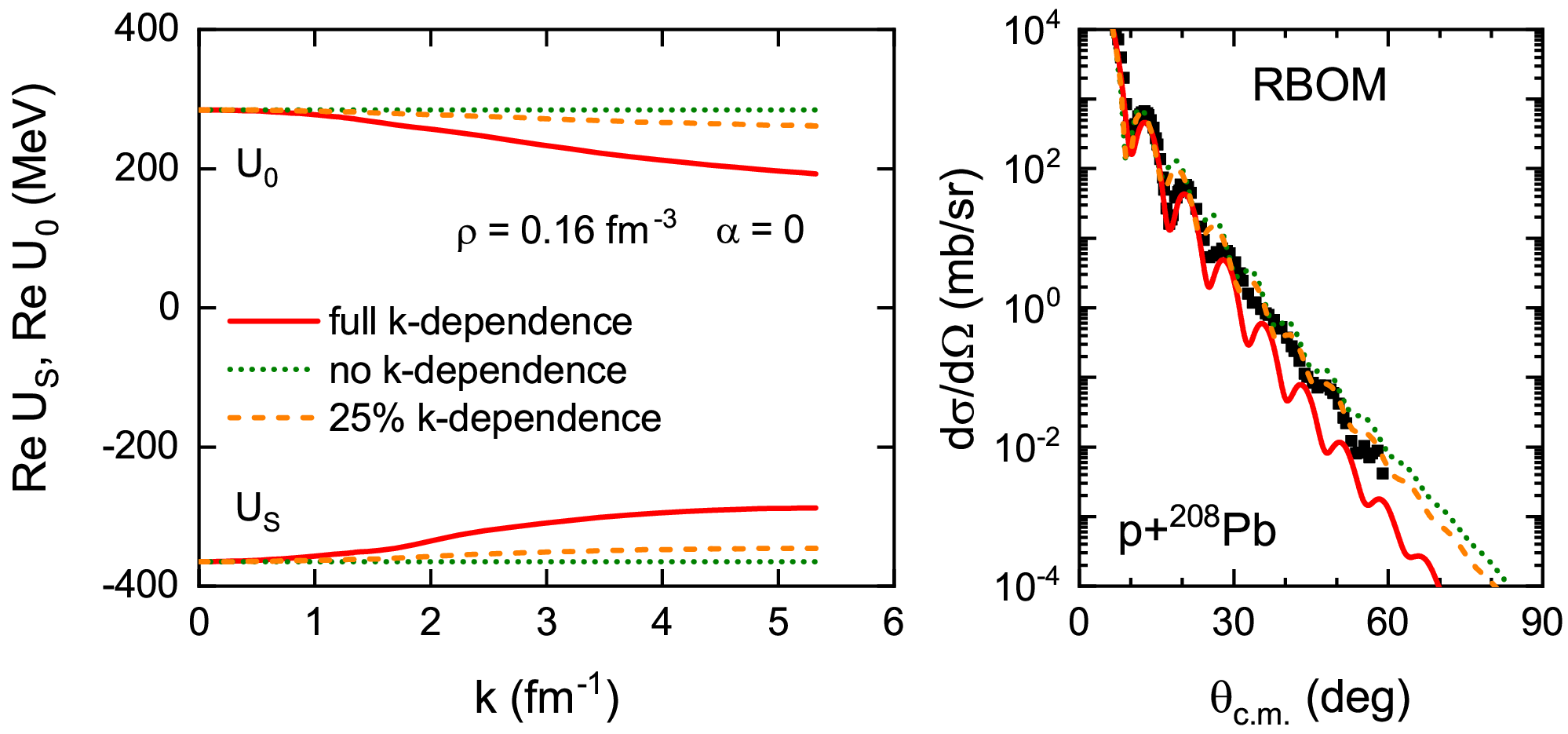}
  \caption{(Color online) Left: The real parts of single-particle potentials $U_S$ and $U_0$ as a function of momentum for symmetric nuclear matter at empirical saturation density. From solid, short-dashed, to short-dotted, the momentum dependences of Re$U_S$ and Re$U_0$ are decreasing. Right: The corresponding cross section angular distributions for $p+\prescript{208}{}{\text{Pb}}$ with incident energy 200 MeV.}
  \label{label-FigS1}
\end{figure}

In the left panel of Fig.~\ref{label-FigS1}, we show the real part of single-particle potentials $U_S$ and $U_0$ as a function of momentum $k$ in symmetric nuclear matter. The red solid line labeled "full $k$-dependence" corresponds to RBOM itself. At high momentum, the magnitudes of both potentials are weakened, exhibiting a somewhat large momentum dependence. To investigate the influence of the momentum dependence of $U_S$ and $U_0$ on the cross section angular distributions, we make two modifications. The one labeled "no $k$-dependence" is obtained by eliminating the momentum dependence by forcing $\text{Re}U_i(k)=\text{Re}U_i(0)$ with $i=S,0$, while the other one labeled "$25\%$ $k$-dependence" is obtained by replacing $\text{Re}U_i(k)$ by $\text{Re}U_i(k)\times0.25+\text{Re}U_i(0)\times0.75$. The corresponding cross section angular distributions for $p+\prescript{208}{}{\text{Pb}}$ with incident energy 200 MeV are shown in the right panel of Fig.~\ref{label-FigS1}. It is clear that, by decreasing the momentum dependence of Re$U_S$ and Re$U_0$, larger cross sections and better agreement with experimental data can be obtained, especially for the modification that keeps only 25\% momentum dependence. This indicates that the large momentum dependence of single-particle potentials in nuclear matter is responsible for the underestimation of the cross section angular distributions at high incident energies.

\begin{figure}[htbp]
  \centering
  \includegraphics[width=8.0cm]{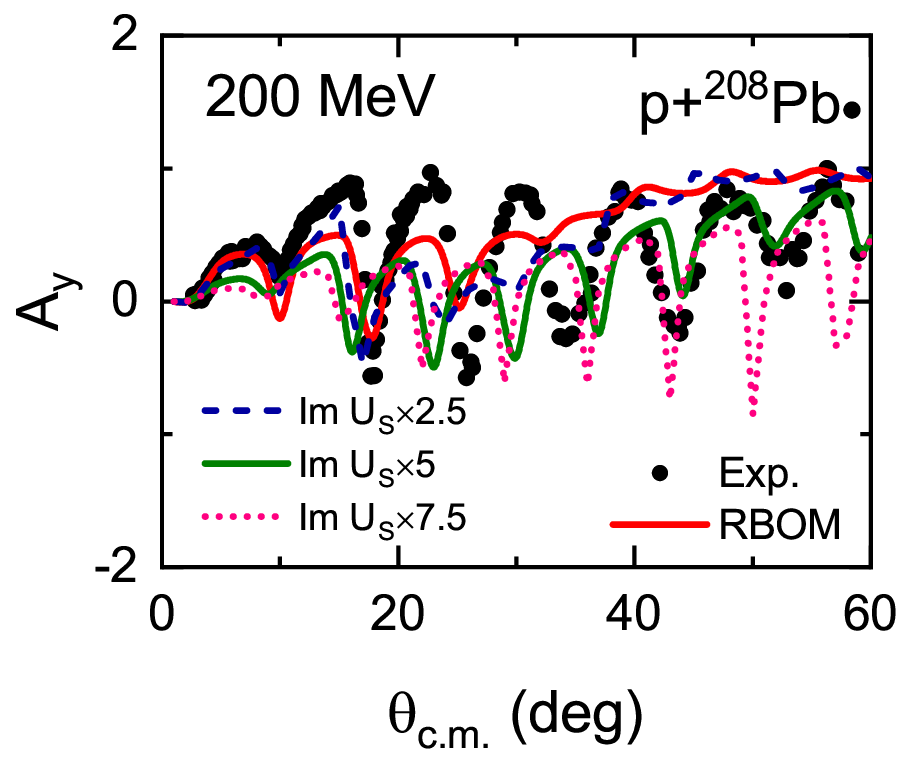}
  \caption{(Color online) The analyzing power $A_y$ for $p+\prescript{208}{}{\text{Pb}}$ with incident energy 200 MeV calculated by the RBOM potential, with the magnitudes of Im$U_S$ being increased by factors of 2.5, 5, and 7.5. Notice that the momentum dependence of Re$U_i$ with $i=S,0$ has been modified as shown in Fig.~\ref{label-FigS1}.}
  \label{label-FigS2}
\end{figure}

For the analyzing power $A_y$ which is dominated by the spin-orbit potentials, firstly we find a limited effects on $A_y$ at large angles by doubling Re$V_{\text{so}}$ in Fig.~\ref{label-FigS0}. Therefore, the deviation for $A_y$ is not mainly affected by the real part of spin-orbit potentials.
Considering that the magnitude of Im$U_S$ is smaller than that of Im$U_0$ as can be seen in Figs.~\ref{label-Fig1} and \ref{label-Fig2}, we then manually increase the magnitude of Im$U_S$ to increase Im$V_{\text{so}}$ and to see the effects on $A_y$. In Fig.~\ref{label-FigS2}, it is found that increasing the magnitude of Im$U_S$ by a factor of 5 does improve the prediction on $A_y$, especially for large angles. This indicates that the small magnitude of the imaginary part of $U_S$ is related to the unsatisfactory theoretical predictions of the analyzing power $A_y$ at high incident energies and large angles.

\subsection{Uncertainty quantification for the RBOM potential}


\begin{figure}[htbp]
  \centering
  \includegraphics[width=16.0cm]{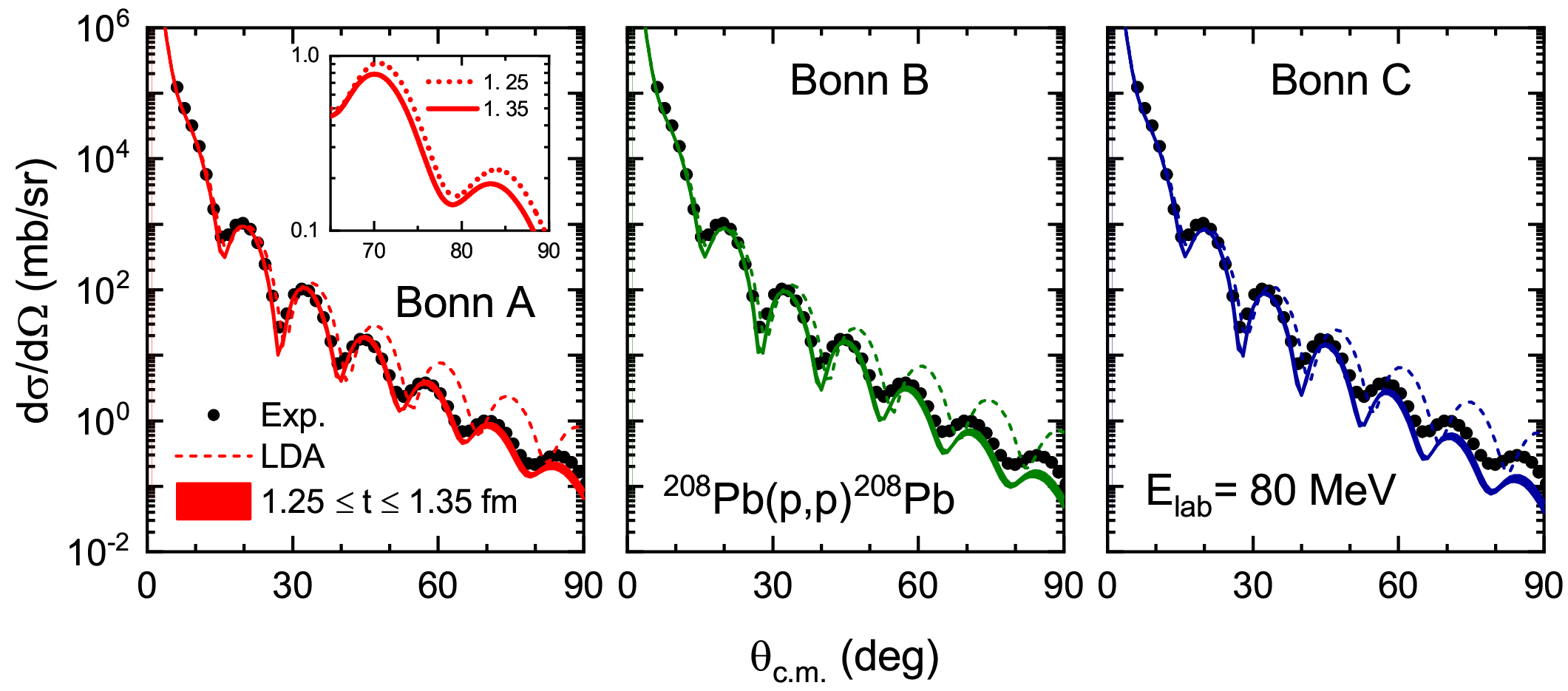}
  \caption{(Color online) The elastic scattering differential cross section as a function of scattering angles in the c.m. frame, calculated by the RBOM potential for proton scattering off $\prescript{208}{}{\text{Pb}}$ with incident energy 80 MeV. 
  The left, middle, and right panels correspond to Bonn A, B, and C respectively.
  In each panel, the results from ILDA with parameter $t$ ranging from 1.25 to 1.35 fm (shaded regions) as well as LDA (dashed lines) are shown.
  The black dots are experimental data.
  Results with $t=1.25$ and 1.35 fm with the Bonn A potential are shown as the inset in the left panel.}
  \label{label-Fig17}
\end{figure}

The uncertainty quantification of optical potentials has attracted increasing attention~\cite{2019-King-PhysRevLett.122.232502, 2022-Baker-PhysRevC.106.064605, 2023-Pruitt-PhysRevC.107.014602}.
The RBOM potential is developed based on the RBHF theory with the improved local density approximation, where the realistic $NN$ interaction Bonn A is used.
Therefore, the uncertainty of the RBOM potential has two sources.
On the one hand, the effective range parameter $t$ in the ILDA can not be derived from the RBHF calculation. 
The relevant uncertainty can be analyzed by varying the range parameter $t$ within a suitable range.
On the other hand, the $NN$ scattering data and two-body properties are not sufficient to constrain the parameters of realistic $NN$ interactions.
Even though there is no such a systematic expansion framework for the one-meson-exchange model, the uncertainty from the $NN$ interaction adopted in this work can be analyzed by using three different parametrizations of the Bonn potential, i.e., Bonn A, B, and C.
In the left panel in Fig.~\ref{label-Fig17}, we show the elastic scattering differential cross section for proton scattering on $\prescript{208}{}{\text{Pb}}$ calculated with the parameter $t$ ranging from 1.25 to 1.35 fm.
It is found that the uncertainty due to the adjustable parameters of the ILDA is very small, especially for small scattering angles.
The inset in this panel shows that increasing the parameter $t$ reduces the elastic scattering differential cross section, and the reduction is more apparent around the maxima. 
The case with LDA is also depicted in this panel and large derivations from experimental data are found.
The corresponding results with Bonn B and Bonn C are shown in the middle and right panels in Fig.~\ref{label-Fig17}, 
The difference among the three potentials are mainly reflected at large angles, where the performance of Bonn A is the best for this incident energy.


\section{Summary}\label{sec:summary}

In summary, the RBOM potential, i.e., the relativistic microscopic $O$ptical $M$odel potential for nucleon-nucleus scattering based on the $R$elativistic $B$rueckner-Hartree-Fock (RBHF) theory in combination with the improved local density approximation (ILDA), has been developed.
The RBHF calculations for symmetric and asymmetric nuclear matter are performed in the full Dirac space with realistic nucleon-nucleon ($NN$) interaction chosen as Bonn A. 
The full-Dirac-space calculations have determined the single-particle potentials uniquely by considering the positive-energy and negative-energy states simultaneously, thus avoiding the usually used approximations due to neglecting negative-energy states.
The density distributions of target nuclei are calculated with the relativistic density functional PC-PK1.   
The single-particle potentials at low density below 0.08 $\text{fm}^{-3}$ are extrapolated by quadratic functions. 
Except for the effective range parameter in the ILDA, there is no free parameter in the RBOM potential. 

Overall, the RBOM potential reproduces the elastic scattering differential cross sections for stable targets $\prescript{208}{}{\text{Pb}}$, $\prescript{120}{}{\text{Sn}}$, $\prescript{90}{}{\text{Zr}}$, $\prescript{48}{}{\text{Ca}}$, and $\prescript{40}{}{\text{Ca}}$ with incident energy below 200 MeV.
The results of the optical potential is comparable to the widely used phenomenological relativistic global optical potential. 
The predictions of the analyzing power $A_y$, spin rotation function $Q$, and reaction cross section $\sigma_{\text{reac}}$ for proton scattering off $\prescript{208}{}{\text{Pb}}$ are also found to be consistent with the experimental data.
The uncertainties of the optical potential resulting from the range parameter in ILDA and $NN$ interactions are examined and found to be minor.

The description of the neutron-nucleus scattering will be presented in a forthcoming paper.
For further evaluation of the performance of the RBOM potential near stability, systemic studies in a wide range of mass numbers $12\leq A\leq 209$ and incident energies $E\leq 200$ MeV are necessary.
In parallel, applying the RBOM potential to exotic nuclei will provide a reliable framework to investigate the isospin effects in nuclear structure from a scattering prospect.
Furthermore, by folding the $G$ matrix in coordinate space with the target densities, one can go beyond the ILDA and construct a relativistic microscopic optical potential in a more advanced way.
It is interesting to investigate whether the optical potential from $G$ matrix folding has positive impact on the improvement of current descriptions at large incident energies and scattering angles.

\begin{acknowledgments}

S.W. thanks R. Xu for helpful discussions and E.D. Cooper for providing the newest code of GOP.
This work was supported in part by the National Natural Science Foundation of China (NSFC) under Grants No. 12205030, No. 12347101, and No. 12375126, the Fundamental Research Funds for the Central Universities under Grants No. 2020CDJQY-Z003 and No. 2021CDJZYJH-003, the Institute for Basic Science under Grant No. IBS-R031-D1, and by the Deutsche Forschungsgemeinschaft (DFG, German Research Foundation) under Germanys Excellence Strategy EXC-2094-390783311, ORIGINS.
Part of this work was achieved by using the supercomputer OCTOPUS at the Cybermedia Center, Osaka University under the support of the Research Center for Nuclear Physics of Osaka University and the High Performance Computing Resources in the Research Solution Center, Institute for Basic Science.

\end{acknowledgments}

\bibliography{paper-NASca}

\end{document}